\newif\ifjournal\journalfalse
\newif\ifarxiv\arxivtrue
\ifjournal
\documentclass[twoside,twocolumn,9pt]{article}
\fi
\ifarxiv
\documentclass[twocolumn,9pt]{article}
\fi

\usepackage{extsizes}
\usepackage{notes2bib}
\usepackage[super,sort&compress,comma]{natbib} 
\usepackage[version=3]{mhchem}

\usepackage[left=1.5cm, right=1.5cm, top=1.785cm, bottom=2.0cm]{geometry}

\usepackage{balance}
\usepackage{mathptmx}
\usepackage{sectsty}
\usepackage{graphicx} 
\usepackage[utf8]{inputenc}
\usepackage{lastpage}
\usepackage[format=plain,justification=justified,singlelinecheck=false,font={stretch=1.125,small,sf},labelfont=bf,labelsep=space]{caption}
\usepackage{float}
\usepackage{fancyhdr}
\usepackage{amsmath}
\usepackage{comment}
\usepackage{fnpos}
\usepackage[english]{babel}
\addto{\captionsenglish}{%
	
}
\usepackage{array} 
\usepackage{droidsans}
\usepackage{charter}
\usepackage[T1]{fontenc}
\usepackage[usenames,dvipsnames]{xcolor}
\usepackage{setspace}
\usepackage[compact]{titlesec}
\usepackage[hidelinks]{hyperref}
\usepackage{bbold}

\usepackage{braket}

\title{\Huge \sffamily\textbf{Diamond surface engineering for molecular sensing\\ with nitrogen--vacancy centers\\$\,$}}
\author{\sffamily{\Large Erika Janitz, Konstantin Herb, Laura A. V\"{o}lker, William S. Huxter, Christian L. Degen} \\  \sffamily{\Large and John M. Abendroth}}
\date{}

\usepackage{epstopdf}%

\definecolor{cream}{RGB}{222,217,201}

\begin{document}
	
	\pagestyle{fancy}
	\thispagestyle{plain}
	\fancypagestyle{plain}{
		\renewcommand{\headrulewidth}{0pt}
	}
	
	\makeFNbottom
	\makeatletter
	\renewcommand\LARGE{\@setfontsize\LARGE{15pt}{17}}
	\renewcommand\Large{\@setfontsize\Large{12pt}{14}}
	\renewcommand\large{\@setfontsize\large{10pt}{12}}
	\renewcommand\footnotesize{\@setfontsize\footnotesize{7pt}{10}}
	\makeatother
	
	\renewcommand{\thefootnote}{\fnsymbol{footnote}}
	\renewcommand\footnoterule{\vspace*{1pt}%
		\color{cream}\hrule width 3.5in height 0.4pt \color{black}\vspace*{5pt}} 
	\setcounter{secnumdepth}{5}
	
	\makeatletter 
	\renewcommand\@biblabel[1]{#1}            
	\renewcommand\@makefntext[1]%
	{\noindent\makebox[0pt][r]{\@thefnmark\,}#1}
	\makeatother 
	\renewcommand{\figurename}{\small{Fig.}~}
	\sectionfont{\sffamily\Large}
	\subsectionfont{\normalsize}
	\subsubsectionfont{\bf}
	\setstretch{1.125} %
	\setlength{\skip\footins}{0.8cm}
	\setlength{\footnotesep}{0.25cm}
	\setlength{\jot}{10pt}
	\titlespacing*{\section}{0pt}{4pt}{4pt}
	\titlespacing*{\subsection}{0pt}{15pt}{1pt}
	
	\fancyfoot{}
	\fancyfoot[LO,RE]{\vspace{-7.1pt}\includegraphics[height=9pt]{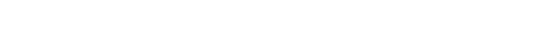}}
	\fancyfoot[RO]{\footnotesize{\sffamily{\thepage}}}
	\fancyfoot[LE]{\footnotesize{\sffamily{\thepage}}}
	\fancyhead{}
	\renewcommand{\headrulewidth}{0pt} 
	\renewcommand{\footrulewidth}{0pt}
	\setlength{\arrayrulewidth}{1pt}
	\setlength{\columnsep}{6.5mm}
	\setlength\bibsep{1pt}
	
	\makeatletter 
	\newlength{\figrulesep} 
	\setlength{\figrulesep}{0.5\textfloatsep} 
	
	\newcommand{\topfigrule}{\vspace*{-1pt}%
		\noindent{\color{cream}\rule[-\figrulesep]{\columnwidth}{1.5pt}} }
	
	\newcommand{\botfigrule}{\vspace*{-2pt}%
		\noindent{\color{cream}\rule[\figrulesep]{\columnwidth}{1.5pt}} }
	
	\newcommand{\dblfigrule}{\vspace*{-1pt}%
		\noindent{\color{cream}\rule[-\figrulesep]{\textwidth}{1.5pt}} }
	
	\makeatother
	
	\twocolumn[
	\begin{@twocolumnfalse}

		\ifjournal
		\begin{tabular}{m{4.5cm} p{13.5cm} }
			\fi

			\ifarxiv
			\vspace{1cm}
			\maketitle
			\vspace{-1cm}
			\fi

			\ifarxiv
			\begin{center}\vspace{0.5cm}{\sffamily{Department of Physics, ETH Z\"{u}rich, Otto-Stern-Weg 1, 8093 Z\"{u}rich, Switzerland}} \end{center}\vspace{0.5cm}
			\fi

			{
				\ifarxiv
				\begin{center}\large\begin{minipage}{16cm}
						\fi
						Quantum sensing using optically addressable atomic-scale defects, such as the nitrogen--vacancy (NV) center in diamond, provides new opportunities for sensitive and highly localized characterization of chemical functionality. Notably, near-surface defects facilitate detection of the minute magnetic fields generated by nuclear or electron spins outside of the diamond crystal, such as those in chemisorbed and physisorbed molecules. 
						However, the promise of NV centers is hindered by a severe degradation of critical sensor properties, namely charge stability and spin coherence, near surfaces (< \textit{ca.} 10 nm deep). Moreover, applications in the chemical sciences
						require methods for covalent bonding of target molecules to diamond with robust control over density, orientation, and binding configuration.
						This forward-looking Review provides a survey of the rapidly converging fields of diamond surface science and NV-center physics, highlighting their combined potential for quantum sensing of molecules. 
						We outline the diamond surface properties that are advantageous for NV-sensing applications, and discuss strategies to mitigate deleterious effects while simultaneously providing avenues for chemical attachment. Finally, we present an outlook on emerging applications in which the unprecedented sensitivity and spatial resolution of NV-based sensing could provide unique insight into chemically functionalized surfaces at the single-molecule level.
						\ifarxiv \hspace{0.1cm}\\ \begin{center}\includegraphics[width=11cm]{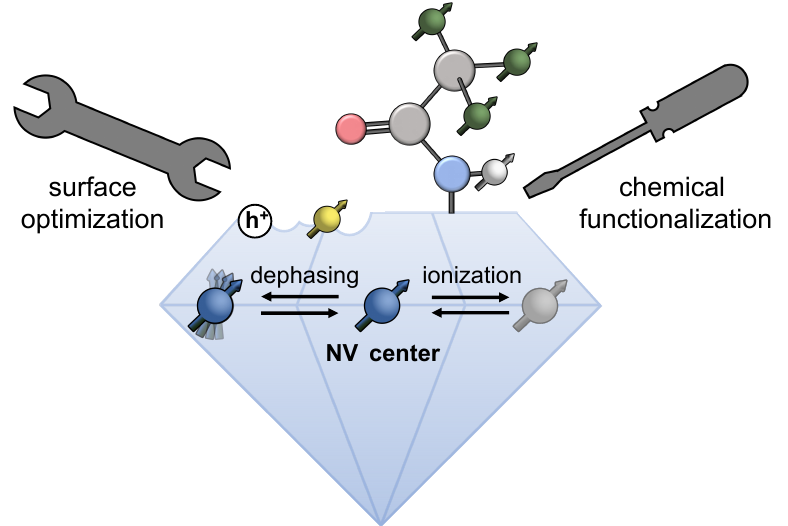} \vspace{8cm}\end{center}\end{minipage}\end{center}\fi
			} %

			\ifjournal
		\end{tabular}
		\fi
	\end{@twocolumnfalse} \vspace{0.6cm}
	
	]
	
	\renewcommand*\rmdefault{bch}\normalfont\upshape
	\rmfamily
	\section*{}
	\vspace{-1cm}

	\section{Introduction}
	Nuclear and electron spins within molecules provide invaluable handles with which to identify chemical structures and intermolecular interactions, allowing for visualization of physiological processes and biological matter for medical diagnostics. Moreover, observing the spin dynamics of transient reaction intermediates could provide new insights into spin-selective chemistries.
	Traditional methods for performing such characterizations include nuclear magnetic resonance (NMR) and electron paramagnetic resonance (EPR) spectroscopy.\cite{Ernst_ACIE_1992,schweiger_Book_2001,Wuthrich_ACIE_2003,Roessler_CSR_2018} Unfortunately, these conventional techniques commonly rely on large spin ensembles to overcome low thermal polarization and suffer from poor detection efficiency, which preclude the study of dilute spin samples.\cite{Jeschke_Jmagres_2004,Terreno_ChemREv_2010,Lee_JMagRes_2014} In addition, spatial resolution is hampered by the challenge of applying large magnetic gradients with nanoscale precision.\cite{Degen_PNAS_2009,Grinolds_NN_2014} Therefore, the grand goal of studying individual molecules (spins) requires a fundamentally different experimental approach.
	
	One such approach replaces traditional NMR and EPR detectors with single, individually addressable quantum spins, which are capable of detecting the minute magnetic fields generated by nuclei or unpaired electrons in nearby molecules.~\cite{Degen_RevModPhys_2017,Wolfowicz_NatRevMater_2021} 
	In particular, the nitrogen--vacancy (NV) center in diamond has emerged as a promising candidate sensor for molecular analysis due to its exquisite magnetic-field sensitivity, nanoscale spatial resolution, biocompatibility, and operational capacity under ambient conditions.\cite{Gruber_Science_1997,Jelezko_PRL_2004,Balasubramanian_Nature_2008,Degen_APL_2008,Doherty_PhysRep_2013,Schirhagl_AnnRevPhysChem_2014,Bucher_NProto_2019,Weggler_NL_2020,Wang_NL_2021} 
	Already in the context of chemical sensing, the atomic scale of such defects has been exploited to probe nanotesla magnetic fluctuations with nanoscale resolution,\cite{Staudacher_Science_2013,Mamin_science_2013} facilitating the study of
	dilute protein assemblies,\cite{Shi_Science_2015,Lovchinsky_Science_2016} DNA,\cite{Shi_natmethods_2018} and paramagnetic species\cite{Sushkov_NL_2014} with single NVs, as well as detecting NMR chemical shifts in proximal molecules.\cite{Aslam_Science_2017,Glenn_Nature_2018,Arunkumar_PRXQuant_2021}
	However, the weak dipolar fields generated by target spins decay rapidly with target--sensor separation, motivating the use of near-surface NVs and direct molecular functionalization of the diamond surface.
	In such cases, the exquisite sensitivity of the NV center presents a challenge in the presence of surface noise, which degrades charge stability and spin coherence for shallow defects.
	\cite{Sangtawesin_PRX_2019,Bluvstein_PRL_2019b}
	Thus, improving near-surface NV properties while simultaneously enabling chemical functionalization of the diamond surface represents a
	critical multidisciplinary challenge.

	This Review begins with
	a tutorial-style introduction to the NV center electronic structure and sensing properties, focusing on detection of magnetic fields since it is the most prevalent sensing modality. We include a brief complementary summary of electric-field, strain, and temperature sensing,
	which may prove useful in future surface-chemistry characterization. We follow with a discussion of measurement sensitivity and other relevant figures of merit, elucidating which experimental parameters are most important for chemical characterization with shallow defects. Next, we explore the origins of instabilities that plague near-surface NV centers along with experimental progress toward mitigating these effects. Focus is given mainly to experiments on NV centers hosted in bulk diamond materials as opposed to scanning probe experiments and nanodiamonds, although, we note that many of the surface engineering techniques described here can be applied to those systems. Ultimately, applications in chemical sciences require precise and optimal placement of analytes on the diamond surface; 
	we therefore highlight advances in chemical functionalization techniques that are compatible with near-surface NVs. 
	This discussion is accompanied by a survey of measurement-based strategies for further improving detection sensitivities.
	Finally, we combine the aforementioned chemical toolbox and library of quantum control strategies to
	offer a perspective on untapped applications for NV-based quantum sensing of chemical systems.

	\section{Quantum sensing with the NV center}
	\label{sec:NV}
	
	The NV center is the most widely used and best understood crystallographic defect in diamond.~\cite{Gali_nanophotonics_2019} Its utility as a quantum sensor stems from a number of valuable properties: i) atomic size, which provides nanoscale resolution, ii) energetic coupling to a variety of physical quantities, iii) long spin lifetimes (even at room temperature), iv) coherent spin-state manipulation using microwave or optical fields, and v) spin readout \textit{via} spin-dependent fluorescence. In this section, we will introduce the NV center, discuss its basic sensing properties, and walk through several canonical sensing schemes with the goal of elucidating which sensor parameters are critical in the NV-sensing community. 
	
	\subsection{Physical and electronic structure of the NV center}
	\label{sec:NVstruct}
\begin{figure*}
		\centering
		\includegraphics[width=\textwidth]{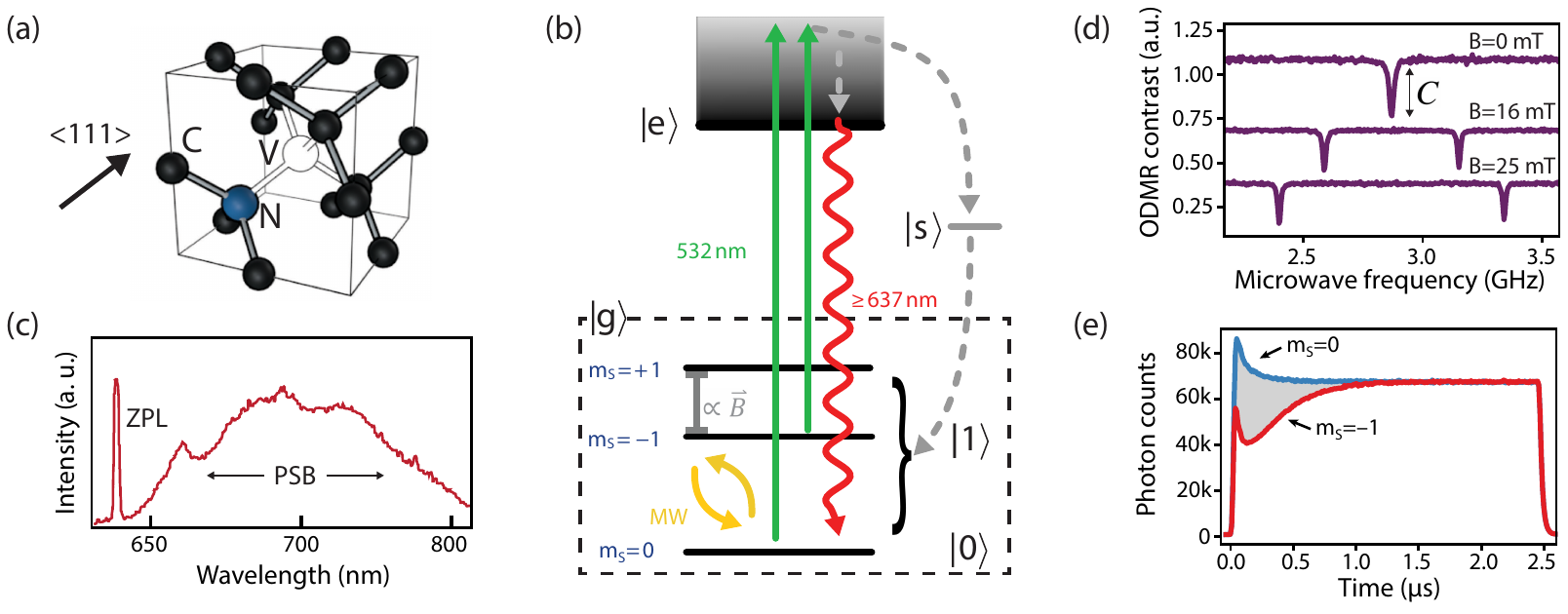}
		\caption{The NV center. (a) Crystallographic structure. (b) Simplified NV electronic structure including ground-state fine structure. (c) NV low-temperature emission spectrum. (d) Continuous-wave optically detected magnetic-resonance (CW ODMR) measurements corresponding to different on-axis DC magnetic fields ($B_0$). Continuous optical and microwave excitation is applied to the NV and the microwave frequency is scanned across ground-state splitting. Dips in fluorescence with contrast $C$ appear when the microwave frequency is resonant with the transitions between $m_s=0$ and $m_s=\pm1$. (e) Transient (time-resolved) spin-dependent fluorescence measurements for an NV center prepared in either the $m_s=0$ or $m_s=-1$ spin states. (c) Adapted from Ref.~\citenum{jelezko_PSSA_2006}, Copyright 2006 Wiley-VCH
			Verlag GmbH \& Co. KGaA}
		\label{fig:NV}
	\end{figure*}

	The NV center comprises a substitutional nitrogen atom and adjacent vacancy occurring along the $\braket{111}$-family of crystallographic directions (Fig. \ref{fig:NV}a).~\cite{Doherty_PhysRep_2013} Due to the diamond crystallographic structure, the NV center has threefold $C_{3v}$ symmetry, where the $z$-axis is typically defined along the nitrogen--vacancy bond and the $x$-axis points orthogonally towards one of the three carbons closest to the vacancy. In its neutral state, five electrons contribute to the net electronic spin; one from each of the three carbon dangling bonds and two from the nitrogen lone pair. When negatively charged by an additional electron, it develops a $S = 1$ spin character, which is essential for its use as a quantum sensor.~\cite{Gali_PRB_2008,Maze_NJP_2011,Doherty_PRB_2012} 
	
	The relevant low-energy states of the NV lie within the band gap of diamond, hosting an electronic ground-state $\left(\ket{g}\right)$ with orbital-singlet, spin-triplet character (Fig.~\ref{fig:NV}b). This state couples to the environment according to the ground-state Hamiltonian,~\cite{Doherty_PRB_2012,Doherty_PhysRep_2013} which can be written as
	\begin{multline}
		\mathcal{H}_\mathrm{gs}=  
		\underbrace{h\, D_\mathrm{gs}\, \left(\hat{S}_z^2 -\frac{2}{3} \mathbb{1}\right)}_{\text{ZFS}}
		+ \underbrace{\hbar\, \gamma_\mathrm{NV}\, \vec{S}\cdot \vec{B}}_{\text{Zeeman}} 
		+\underbrace{h\sum_{i=1}^{M}\vec{S} {\bf A}_i(\vec{r}_i)\vec{I}_i}_{\text{HFI}} \\
		+ \underbrace{h\, d_\mathrm{z}\, \mathcal{E}_\mathrm{z}\, \left(\hat{S}_z^2 -\frac{2}{3} \mathbb{1}\right)\, + h\, d_\perp\, \left(\mathcal{E}_\mathrm{x} (\hat{S}_y^2\, - \hat{S}_x^2)\, + \mathcal{E}_\mathrm{y}\, (\hat{S}_x\,\hat{S}_y + \hat{S}_y\,\hat{S}_x ) \right) }_{\text{Stark}}.
		\label{eq:Hgs}
	\end{multline}
	Here, $h$ ($\hbar$) is the (reduced) Planck constant,  $\vec{S} = [\hat{S}_x, \hat{S}_y, \hat{S}_z]$ are the spin-1 operators, $\vec{I}$ is the nuclear spin operator and $\mathbb{1}$ is the identity operator. The simplified Hamiltonian in Eq.~\ref{eq:Hgs} is spanned by the three $S = 1$ spin states: $|m_s=0\rangle$,  $|m_s=+1\rangle$, and $|m_s=-1\rangle$, and is grouped into four terms. First, there is a zero-field splitting (ZFS) term caused by electron spin--spin interaction, which shifts the $|m_s = \pm1\rangle$ states in energy by $D_{\mathrm{gs}}=2.87$ GHz relative to $|m_s=0\rangle$. The second term describes the Zeeman interaction  for the $|m_s = \pm1\rangle$ states, which is used to lift their degeneracy in quantum sensing experiments (Sec.~\ref{sec:Magsens}); most commonly, the anisotropy of the interaction is neglected and the energy splitting is given by the product of the gyromagnetic ratio of the NV center $\gamma_\mathrm{NV}$ and the strength of the magnetic field. Here, the gyromagnetic ratio is $\gamma_\mathrm{NV} = 2 \pi \times \gamma_\mathrm{e}\, (1 + 357\,\mathrm{ppm}) = 2 \pi \times 28.0345(28)\,\mathrm{GHz/T}$\cite{codata_2018,Felton_2009} differing only slightly from the free electron value as typical for carbon-based materials.\cite{Herb_2018} The third term of $\mathcal{H}_\mathrm{gs}$ includes coherent coupling to $M$ proximal nuclear spins \textit{via} the hyperfine interaction (HFI). The hyperfine tensor ${\bf A}_i(\vec{r}_i)$ comprises a dipolar component and a contact contribution for nearby nuclear spins. Finally, the fourth term describes the coupling to both electric fields $\vec{E}$ and strain $\vec{\sigma}$,  ($\vec{\mathcal{E}} = \vec{E} + \vec{\sigma}$), which impact the NV center through electric dipole interaction, piezoelectric coupling, and spin--spin interaction that distort the electron orbitals.~\cite{Maze_NJP_2011,Doherty_PRB_2012}
	Consequently, the three-dimensional structure (and $C_{3v}$ symmetry) of the NV necessitates a directional dependence on
	$\vec{\mathcal{E}}$. The coupling is mediated via the ground-state electric susceptibility parameters of the NV center $d_z = 3.5 \times 10^{-3}$ Hz/(V/m)~\cite{VanOort_ChemPhyLett_1990} and $d_\perp = 0.165$ Hz/(V/m)~\cite{Michl_NanoLett_2019}.  An additional perpendicular electric term (proportional to the coupling parameter $d_\perp^{'}$) is omitted from Eq.~\ref{eq:Hgs}, as it can be ignored when $d_\perp^{'} \sqrt{\mathcal{E}_\mathrm{x}^2 +\mathcal{E}_\mathrm{y}^2} \ll D_{\mathrm{gs}}$,~\cite{Doherty_PRB_2012} which is the case in most experiments.
	
	In addition, the NV center is optically addressable, with an excited state $\left(\ket{e} \text{, Fig.~\ref{fig:NV}b}\right)$ located $1.945$ eV ($637$ nm) above the ground state. This state has spin-triplet, orbital-doublet character (see other reviews for a detailed description~\cite{Jelezko_PhysStatSol_2006,Doherty_PhysRep_2013,Gali_nanophotonics_2019,Barry_RevModPhys_2020}) resulting in a fine structure containing six energy levels, which can be resolved under cryogenic conditions ($T < 10$ K).\cite{Fu_PRL_2009} At elevated temperatures, these orbitals undergo rapid averaging caused by the dynamic Jahn-Teller effect,\cite{Fu_PRL_2009,Happacher_arxiv_2021} resulting in an effective three-level (spin-triplet) system at room temperature. In either case, decay from the NV excited state exhibits spin-dependent fluorescence (discussed in Sec.~\ref{sec:senseprop}) that is exploited for spin-state readout in most NV-based quantum sensing schemes.
	
	\subsection{Photophysics and sensing properties}
	\label{sec:senseprop}
	An ideal quantum sensor would offer mechanisms for coherent control of the sensor state, as well as methods for efficient state preparation and readout. Such a system can be realized within the NV ground state using the $m_s = 0$ and one of the $m_s = \pm1$ states (hereafter generalized as $\ket{0}$ and $\ket{1}$, Fig.~\ref{fig:NV}b) by applying a magnetic field along the $z$-axis to split the $m_s = \pm1$ energy levels (see Fig.~\ref{fig:NV}d). Impressively, resonant microwave fields have been used to achieve universal quantum control within the ground-state manifold~\cite{Jelezko_PRL_2004} with fidelities exceeding $99\%$~\cite{Rong_ncomm_2015} 
	
	Furthermore, the NV center optical transitions can be leveraged for initializing and reading out the spin state.~\cite{Jelezko_PRL_2004,Dreau_PRB_2011}
	Excitation to the excited state is typically achieved using an off-resonant laser (usually $\sim532$ nm, see green arrows in Fig.~\ref{fig:NV}b). The resulting NV fluorescence spectrum (radiative decay from the excited state to the ground state) comprises a zero phonon line (ZPL, $3\%$ of emission)~\cite{Barclay_PRX_2011} and a broad phonon sideband (PSB, $97\%$ of emission) that extends to $800$ nm~\cite{Rondin_PRB_2010} (see red arrow in Figs.~\ref{fig:NV}b and c
	). Such off-resonant schemes allow for spin preparation and readout using an intermediate singlet state $\left(\ket{s}\text{, see Fig. \ref{fig:NV}b}\right)$: $m_s = \pm1$ spin projections are more likely to decay non-radiatively from the excited state into $\ket{s}$, from which there is a roughly equal probability to decay (non-radiatively) to any spin projection in the ground state (grey arrow in Fig.~\ref{fig:NV}b).\cite{Robledo_NJP_2011,Thiering_PRB_2018} Consequently, continuous $532$ nm optical illumination results in a net ground-state spin polarization into $m_s=0$.\cite{Robledo_NJP_2011} Such spin polarization is illustrated in Fig.~\ref{fig:NV}e, where both spin states reach the same fluorescence levels after $\sim 1\ \mu$s of laser excitation. Moreover, this nonradiative decay process provides a readout mechanism for the spin state since it yields reduced fluorescence for $m_s\pm1$ states with up to $\sim40\%$ contrast (Figs. \ref{fig:NV}d-e).\bibnote{The authors have experimentally observed several NV centers with $C$>0.4 throughout their collective work.} The simplest experiment demonstrating this phenomenon is continuous-wave optically detected magnetic resonance (CW ODMR), in which microwave and laser excitation are applied simultaneously. When the microwave frequency is resonant the splittings between $\ket{m_s=0}$ and $\ket{m_s=\pm 1}$, population is transferred and a reduction in fluorescence is observed (Fig.~\ref{fig:NV}d).
	
	The aforementioned preparation, control, and readout mechanisms for the NV offer a powerful toolbox for detection of environmental signals. In practice, sensing applications benefit from several additional experimental constraints: i) proximity to the sensing target for increased signal and spatial resolution, ii) long measurement times for maximizing signal integration and spectral resolution, and iii) efficient mechanisms for spin preparation and readout of the sensor. In the following, we describe relevant sensing properties that are directly influenced by these experimental constraints.
	
	First, we note that the signal strength and detection volume of an NV sensor scales sensitively with experimental geometry, namely with the NV--target distance $d$. Indeed, for high-spatial-resolution detection of rapidly decaying signals, this distance must be minimized through both the NV--interface distance and the interface-target distance. Thus, there has been significant effort toward deterministic fabrication of near-surface NVs for sensing applications, which are outlined in Sec. \ref{sec:nvgen}. Furthermore, diamond surface functionalization methods yielding minimal stand-off distance are discussed in detail in Sec. \ref{sec:func}. 
	
\begin{figure}[h]
		\centering
		\includegraphics[width=0.49\textwidth]{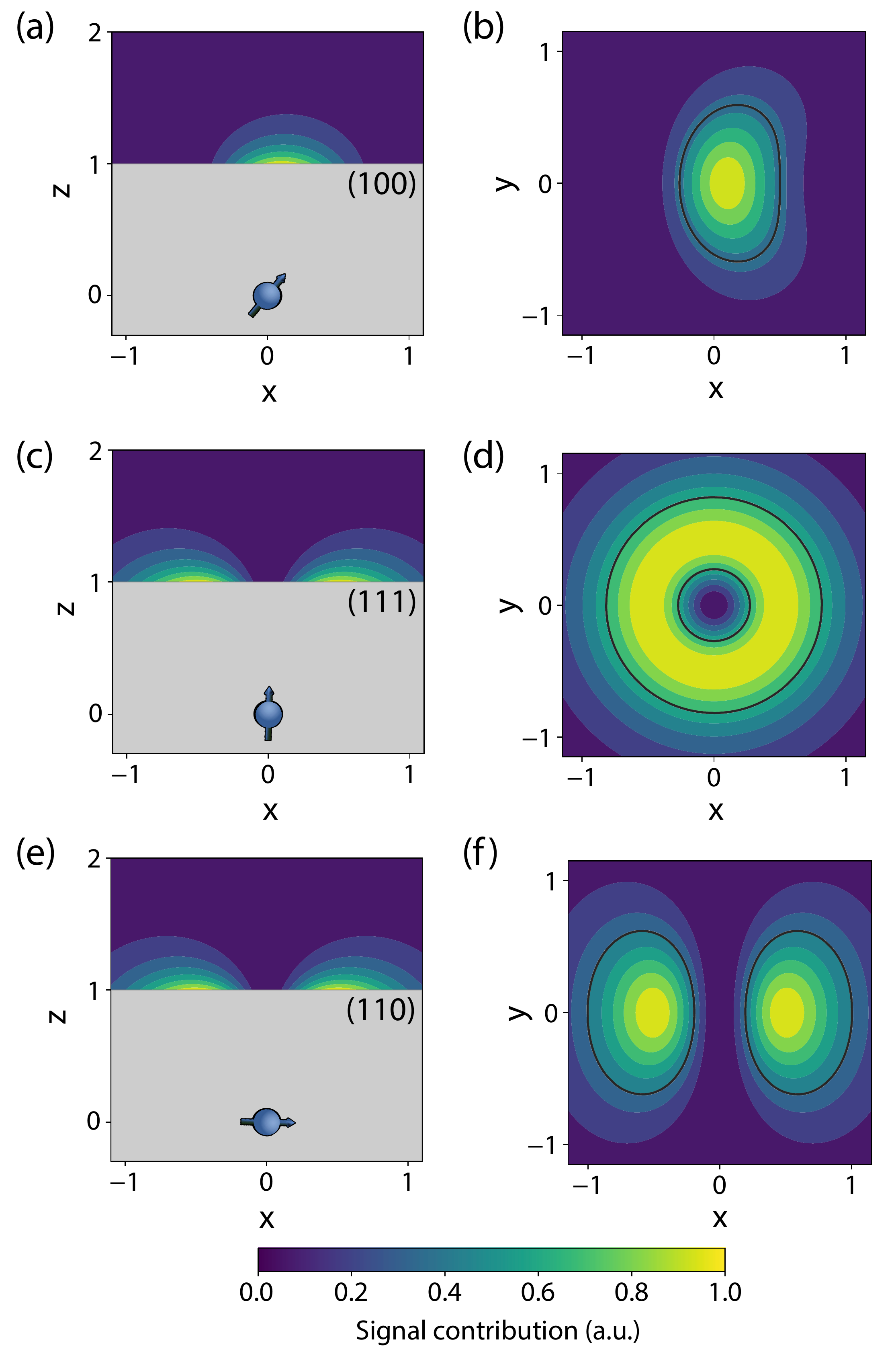}
		\caption{Simulated sensitivity profiles for variance detection of surface spins with different NV orientations. These plots display $B_{\mathrm{rms}}^2$ (in arbitrary units) for an NV at depth $d=1$ below the diamond surface, where all length scales are normalized to this quantity. The areas enclosed with black lines in (b), (d), and (f) contribute $50\%$ of the integrated $B_{\mathrm{rms}}^2$ signal for an infinitesimally-thin 2D layer of surface spins. Note that only fluctuations perpendicular to the NV axis were considered.}
	\label{fig:spatial}
\end{figure}

	The sensor spatial resolution is parameterized by the NV sensing volume, which depends on the specific protocol used and the NV orientation,\cite{Bruckmaier2021} and scales as $V=(0.98 d)^3$ for a $(100)$-cut diamond. Figure \ref{fig:spatial} shows an example sensitivity profile for an echo-like variance detection scheme (protocol details in Sec. \ref{sec:Magsens}). The regions outlined in black contribute $50\%$ of the $B_\mathrm{rms}^2$ signal (or equivalently 70\% of $B_\mathrm{rms}$) for a monolayer of spins on the diamond surface. The specific details of the sensing scheme play an important role, \textit{e.g.,} for phase canceling effects. In general, such sensitivity maps can be obtained by numerically evaluating the signal contribution of a single spin at every location of the diamond surface for a given measurement protocol. In addition, these maps vary strongly with the NV-center crystallographic orientation and cut of the diamond (the surface plane), motivating careful consideration of the substrate.\cite{Bruckmaier2021} 

Next, the measurement time $T_\mathrm{m}$ impacts signal integration and spectral resolution,
and is fundamentally limited by the spin lifetimes of the NV center (either $T_1$, $T_2^*$ or $T_2$ depending on the measurement sequence). The largest of these values is the spin--lattice relaxation lifetime $T_1$ (also known as the longitudinal relaxation lifetime), which characterizes the time for the spin population to reach thermal equilibrium. Impressively, spin--lattice relaxation times within the ground-state manifold can reach a few milliseconds at room temperature~\cite{Jarmola_PRL_2012} and hours at 4 K.~\cite{Abobeih_NatComm_2018} In contrast, $T_2^*$ and $T_2$ refer to the transverse relaxation or spin-coherence lifetimes, which are theoretically limited to $2T_1$ although they are typically much shorter. $T_2^*$ refers to the DC coherence decay time and $T_2$ is used to indicate general, pulsed (AC) coherence decay times. Spin-coherence times of $T_2 > 1$ ms at room-temperature~\cite{Herbschleb_NatComm_2019} and $T_2 \sim 1$ s at $4$ K~\cite{Abobeih_NatComm_2018} have been achieved using multi-pulse experiments (see Sec.~\ref{sec:MeasStrat}).
In practice, low-frequency noise leads to faster dephasing, resulting in $T_2^* \ll T_2$, which limits the ability to measure DC signals.

Furthermore, the spectral resolution $\Delta f$ is an additional consideration for signals that may exhibit closely spaced resonances, such as detection of small chemical shifts, hyperfine couplings, or even spurious harmonics.~\cite{Loretz_PRX_2015,Boss_PRL_2016} For relatively simple sensing protocols, such as those described in the following section, $\Delta f$ is limited by the coherence lifetime of the NV center and sensing protocol ($\Delta f \sim 1/T_2$). However, frequency resolution can be made arbitrarily small~\cite{Boss_Science_2017,Schmitt_Science_2017} and it is possible to decouple the NV coherence time such that resolution is only constrained by the target spin coherence time,~\cite{Cujia_Nature_2019,Pfender_NatureComm_2019} which can be significantly longer (see Sec.~\ref{sec:MeasStrat}).

Next, the magnetic field sensitivity improves for increased spin initialization fidelity. Off-resonant, optical spin polarization efficiency for deep NV centers depends on the applied laser power and can approach $\sim80\%$.~\cite{Robledo_NJP_2011} However, we note that this value can vary considerably for shallow defects due to charge instability.~\cite{Bluvstein_PRL_2019b} Promisingly, logic-based charge initialization (of the negative $\mathrm{NV^-}$ state, see Sec.~\ref{sec:chargestate}) has been employed to increase the average spin initialization fidelity of near-surface defects, with values approaching unity.~\cite{Bluvstein_PRL_2019b} Consequently, we ignore this quantity in subsequent calculations.

Finally, the readout fidelity $F$ quantifies the ability to measure the spin state of the NV at the end of a single experiment, with $F = 1$ corresponding to an ideal, single-shot readout. This value varies considerably for different experimental conditions, and is therefore explored extensively in this text. Since quantum sensing requires repeated measurements for statistical averaging, a larger readout fidelity can speed up experiments significantly. For off-resonant readout, the fidelity is given by~\cite{Taylor_NaturePhys_2008,Hopper_Micromachines_2018}
\begin{align}
	F=\left(1+2\frac{(\alpha_0+\alpha_1)}{(\alpha_0-\alpha_1)^2}\right)^{-1/2},
	\label{eq:F_1}
\end{align}
where $\alpha_0$ and $\alpha_1$ are the expected number of measured photons for $\ket{0}$ and $\ket{1}$, respectively. Here, $\alpha_1=(1-C)\alpha_0$, where $C$ is the measurement contrast. Moreover, $\alpha_0=\xi\,\gamma\,T_\mathrm{R}$, where $\xi$ is the photon collection efficiency, $\gamma$ is the radiative decay rate, and $T_\mathrm{R}$ is the readout time, yielding
\begin{align}
	F=\left(1+4\pi\left(\frac{2-C}{C^2\, \gamma \, \xi \, T_\mathrm{R}}\right)\right)^{-1/2}.
	\label{eq:F_2}
\end{align}
Typical off-resonance fidelities are quite poor ($F \sim 0.03$)~\cite{Taylor_NaturePhys_2008} and much experimental work has been directed at increasing these values (details in Sec.~\ref{sec:ExpCon}).

\subsection{Magnetic sensing}
\label{sec:Magsens}

The vast majority of NV sensing focuses on magnetic interactions owing primarily to its strong coupling constant ($\gamma_\mathrm{NV}$), which impacts detection of external fields as well as the strength of the (dipolar) hyperfine interaction. In the following section, we will explore the sensitivities obtained for some of the simplest detection protocols, including DC and AC schemes with the Ramsey and spin-echo experiments, as well as $T_1$ and $T_2$ relaxation experiments. While these represent only a small fraction of possible sensing schemes, they illustrate which experimental parameters should be considered and optimized when implementing an NV sensor. 

\begin{figure*}
	\centering
	\includegraphics[width=\textwidth]{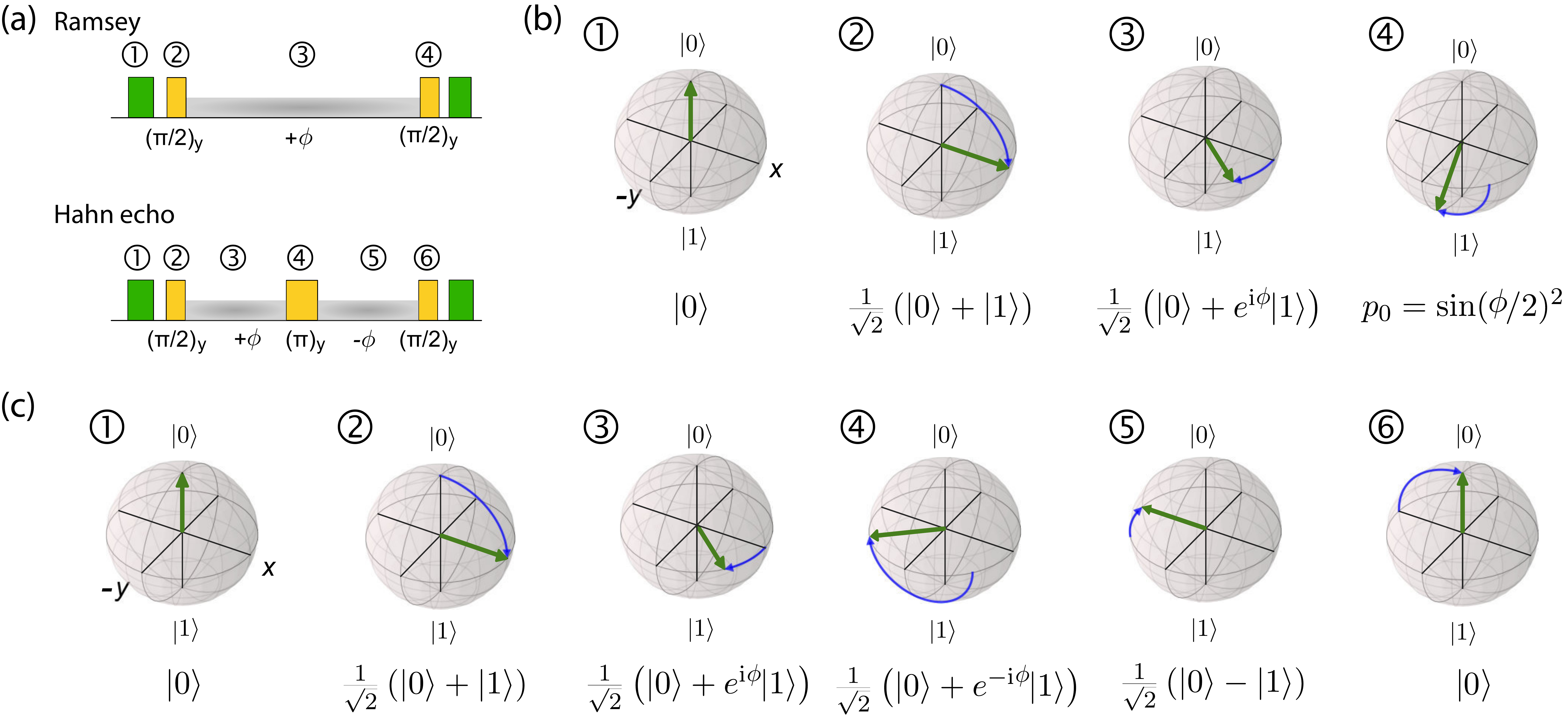}
	\caption{Ramsey and spin-echo sensing experiments. (a) Pulse sequences for the Ramsey and the Hahn-echo protocols. Laser pulses are indicated in green, microwave manipulations are shown in yellow, and phase evolution periods are shaded in grey. (b) Bloch-sphere evolution throughout a Ramsey experiment, (c) Bloch-sphere evolution throughout a spin-echo experiment (steps 1-3 are identical to those of the Ramsey sequence).} 
\label{fig:sens}
\end{figure*}

For these examples, we simplify Eq.~\ref{eq:Hgs} to include only contributions related to a small, time-dependent magnetic field $\vec{B}(t)$ and a relatively large magnetic-field bias $B_0$ along the $z$-axis of the \mbox{NV center}, yielding 
\begin{equation}
\mathcal{H}_\mathrm{gs} = h D_\mathrm{gs}\hat{S}_z^2 + \hbar \gamma_\mathrm{NV} \left(B_0 + B_z(t)\right)\hat{S}_z. \label{eq:eMag}
\end{equation}
Note that Eq. 4 is only accurate in the limiting case where $ \hbar \gamma_\mathrm{NV} B_i(t) \ll h D_{\mathrm{gs}}$ and $B_i(t) \ll B_0$ for $i = \{x, y, z\}$, such that that the NV center becomes insensitive to $B_x(t)$ and $B_y(t)$. The transition energies ($e_{\pm}$) between the $m_\mathrm{s} = 0$ and $m_\mathrm{s} = \pm1$ spin levels are then
\begin{equation}
e_{\pm} = h D_\mathrm{gs} \pm  \hbar \gamma_\mathrm{NV} \left(B_0 + B_z(t)\right).
\label{eq:eMag}
\end{equation}

In this Review, we consider a simplified sensor utilizing a two-level subspace spanned by the $\ket{m_s=0}$ and one of the $\ket{m_s=\pm1}$ states, resulting in the following effective Hamiltonian 
\begin{align}
\label{eq:Hgs2}
\mathcal{H}_\mathrm{gs}= \pm \hbar \gamma_\mathrm{NV} B_z(t)\hat{S}_z.
\end{align}
Here, a rotating frame was used to remove the contributions from $D_\mathrm{gs}$ and $B_0$. We note that an alternative effective Hamiltonian yielding twice the signal can be obtained by instead using the subspace spanned by the two $\ket{m_s\pm1}$ states (details in Sec. \ref{sec:DQS}). The basic sensing premise is as follows: first, the NV state $\ket{\psi}$ is initialized to $\ket{0}$ \textit{via} optical pumping; subsequently, a superposition state $\ket{\psi(t=0)} \propto \ket{0} + \ket{1}$ is created through the application of a $\big(\pi/2\big)_y$ rotation (\textit{e.g.}, with a microwave pulse), which can be visualized on the Bloch sphere~\cite{Feynman_AIP_1957} (Fig. \ref{fig:sens}). While $\ket{0}$ and $\ket{1}$ are eigenstates of Eq.~\ref{eq:Hgs2}, superposition states are not and will therefore evolve in time. In particular, $\ket{0}$ and $\ket{1}$ will acquire a relative phase $\left(\ket{\psi(\tau)} \propto \ket{0} +e^{i\phi(\tau)}\ket{1}\right)$ that is a function of the applied magnetic field
\begin{align}
\label{eq:phi}
\phi(\tau)=\gamma_\mathrm{NV}\int^\tau_0 B_z(t)\,dt,
\end{align}
where $\tau$ is the total evolution time. $\ket{\psi(\tau)}$ will remain in an equal superposition (with equal population in $\ket{0}$ and $\ket{1}$) for all values of $\phi$; however, NV fluorescence can only be used to measure the state populations and not coherences. Application of a second $\big(\pi/2\big)_y$ rotation can convert $\phi$ into a population difference, which can be read out \textit{via} spin-dependent fluorescence. By comparing the resulting fluorescence level against that of the $\ket{0}$ and $\ket{1}$ states, one can infer $\phi$ through the transition probabilities $p_0=\sin^2(\phi/2)$ or $p_1=\cos^2(\phi/2)$, respectively, for the sensing scheme shown in Fig.~\ref{fig:sens}a. Equivalently, $p_i$ is the expectation value of being in state $\ket{i}$ after the last $\big(\pi/2\big)_y$ rotation. %

Typically, one characterizes the magnetic field by measuring the deviation of the transition probability $\delta p=p_i - p$, where $p$ is some fixed bias point. If the measurement protocol can be synchronized with the target signal, the highest sensitivity is obtained for "slope detection",~\cite{Taylor_NaturePhys_2008,Cujia_Nature_2019} wherein $p=0.5$ (corresponding to an equal superposition) or $\phi=k\pi/2$ for odd integers $k$. In the small signal limit ($\delta\phi\ll\pi$), this yields $\delta p=\delta \phi/2$ (for $k=1$). However, many experimental applications preclude synchronization with the signal of interest, resulting in $\langle \delta p \rangle=\langle \delta \phi \rangle=0$ for multiple measurements. In such cases, it is beneficial to instead apply "variance detection",~\cite{Meriles_JCP_2010,Loretz_APL_2014,Pham_PRB_2016} wherein the system is biased at $p=0$ and the variance in transition probability $\langle \delta p^2 \rangle = \frac{1}{4} \langle \delta \phi^2 \rangle=\frac{1}{4}\phi_\mathrm{rms}^2$ is measured.

The $\text{SNR}$ of slope-detection-based experiments is defined as $\text{SNR}=\delta p/\sigma_p$, 
where $\sigma_p=1/(2\sqrt{N}F)$ is a combination of the quantum projection noise for $N$ experimental repetitions and the classical readout noise associated with the finite readout efficiency (described by $F$).~\cite{Degen_RevModPhys_2017}

This analysis can be applied directly in determining the sensitivity of the {\it Ramsey} experiment,~\cite{Ramsey_PR_1950} which is one of the simplest methods for determining a DC magnetic field of the form $B_z(t)=B_{\mathrm{DC}}$. Following Eq.~\ref{eq:phi}, such a field would result in an acquired phase of $\phi(\tau)=\gamma_\mathrm{NV}\, B_\mathrm{DC} \tau$ for an evolution time $\tau$ (Fig.~\ref{fig:sens}a). Subsequent conversion to NV population followed by slope detection yields $\delta p=\gamma_\mathrm{NV} \tau \delta B_\mathrm{DC}/2$ with $\text{SNR} = F\gamma_\mathrm{NV} \delta B_\mathrm{DC}\sqrt{\tau\, T}$ for a total experiment duration $T = N \tau$. The sensitivity can then be calculated as $\eta_\mathrm{DC} \approx 1/(F \gamma_\mathrm{NV} \sqrt{\tau})$, which is optimized by maximizing the acquisition time $\tau$. For the Ramsey experiment, $\tau$ is fundamentally limited to the natural dephasing time of the sensor spin $T_2^*$, resulting in a minimum sensitivity of
\begin{align}
\eta_{\mathrm{DC}} \approx 1/\left(F \gamma_\mathrm{NV} \sqrt{T_2^*}\right).
\end{align}

While Ramsey experiments are ideal for sensing DC fields, they are ill-suited for detection of signals that vary quickly compared to $\tau$ since the acquired phase is averaged away over the course of the measurement. For detection of AC magnetic fields, the canonical {\it spin-echo} experiment~\cite{Hahn_PR_1950} incorporates an intermediate $(\pi)_y$ pulse (Fig. \ref{fig:sens}b) that flips the spin, allowing for an effective reversal of the detected field $B_z(t)$ according to
\begin{align}
\label{eq:echo}
\phi_{\mathrm{echo}}(\tau)=\gamma_\mathrm{NV}\int^{\tau/2}_0 B_z(t)\,dt-\gamma_\mathrm{NV}\int_{\tau/2}^{\tau} B_z(t)\,dt.
\end{align}
We consider a magnetic field with frequency $\omega$ and phase $\theta$ described by $B_z(t)=B_\mathrm{AC}\sin(\omega t +\theta)$; if $\theta$ is known, slope detection can be employed, yielding $\phi_\mathrm{echo}=- \frac{4 B_\mathrm{AC} \gamma_\mathrm{NV}}{\omega}\cos(\omega\tau/2+\theta) \sin^2(\omega\tau/4)$. This quantity is maximized for $\omega=2\pi/\tau$ and $\theta=0$, resulting in $\phi_\mathrm{echo}=2 B_\mathrm{AC} \gamma_\mathrm{NV} \tau /\pi$. The sensitivity can then be calculated as
\begin{align}
\eta_\mathrm{AC} \approx \pi/\left(2F \gamma_\mathrm{NV} \sqrt{T_2}\right).
\end{align}
Here, we set $\tau=T_2$ where $T_2$ is the spin-echo coherence time. This protocol has the added benefit of cancelling low-frequency noise occurring on timescales slower than $\tau$; consequently $T_2$ often far exceeds the natural decoherence time $T_2^*$, resulting in an improvement in slope-detection sensitivity of $\eta_\mathrm{DC}/\eta_\mathrm{AC}\approx\sqrt{T_2/T_2^*}$.

Similarly, the sensitivity for variance detection of an AC field can be calculated from $\phi_\mathrm{rms}=\sqrt{2} B_\mathrm{AC} \gamma_\mathrm{NV} \tau /\pi$ and $\text{SNR} =\langle \delta p^2\rangle / \sigma_p$ resulting in 
\begin{align}
\eta_\mathrm{var} \approx \pi/\left(\gamma_\mathrm{NV} \sqrt{F} \sqrt[^4]{T_2^{3}}\right).
\end{align}

Complementing these widely used DC and AC measurement techniques, {\it relaxometry}~\cite{Schoelkopf_Spinger_2003,Cole_Nanotech_2009} offers a sensing modality for detecting magnetic and electric noise at $\omega\,_{01}= e_\pm /\hbar$, the frequency splitting between $\ket{m_s=0}$ and $\ket{m_s=\pm1}$. While the Ramsey and spin-echo protocols operate in the $0$ Hz (DC) to $< 1$ GHz regimes, $\omega\,_{01}$ may exist in the few GHz regime, extending the frequency range over which NV sensors can be employed.~\cite{Hall_PRL_2009,Hall_NatComms_2016}

In such experiments, the system is initialized to $\ket{0}$ and allowed to evolve for time $\tau$. Subsequently, $\delta p$ is measured, displaying an exponential decay as a function of time with rate
\begin{align}
\Gamma_{1}=\frac{1}{T_1}=\frac{\gamma_{\mathrm{NV}}^2}{2}S_{B,\perp}(\omega\,_{01}).
\end{align}
Here, $S_{B,\perp}(\omega)$ is the power spectral density of the transverse magnetic field,~\cite{Degen_RevModPhys_2017} which appears since perpendicular magnetic fields induce spin flips resulting in $T_1$ relaxation. In addition, relaxometry involving high-frequency noise related to electric fields has also been demonstrated.~\cite{Kim_PRL_2015,Li_PRL_2020} 
If the system is instead prepared in a superposition state $\ket{\psi(t=0)} \propto \ket{0} +\ket{1}$ then the associated decay time also depends on the parallel component of the noise spectral density $S_{B,z}(\omega)$, yielding
\begin{align}
\Gamma_2&=\frac{1}{2 T_1}+\frac{1}{T_2^*}
=\frac{\gamma_\mathrm{NV}^2}{4}S_{B,\perp}(\omega\,_{01})+\frac{\gamma_\mathrm{NV}^2}{2}S_{B,z}(0),
\label{eq:gamma2}
\end{align}
where $S_{B,z}$ is probed at $\omega=0$ since phase flips do not require energy.~\cite{Degen_RevModPhys_2017} Typically, $T_1 \gg T_2^*$ such that $\Gamma_2 \approx 1/T_2^*$, yielding the natural dephasing time observed in Ramsey experiments. Indeed, this decay rate can be thought of in terms of the rms phase previously mentioned in the context of variance detection, where $\phi^2_\mathrm{rms}=2\Gamma \tau$. Moreover, extension of this technique with multi-pulse sequences (such as spin-echo or the dynamical decoupling sequences described in Sec.~\ref{sec:MeasStrat}) enable measurements of $S_{B,z}$ for non-zero $\omega$.~\cite{Steinert_NatComms_2013,Romach_PRL_2015}

\subsection{Electric sensing}
\label{sec:Elecsens}

Electric-field sensing with NV centers is possible through a piezoelectric coupling that produces a Stark shift in the NV resonance levels.~\cite{Tamarat_PRL_2006,Maze_NJP_2011,Doherty_PRB_2012,Gali_nanophotonics_2019} While such schemes suffer from poor sensitivities (in comparison to magnetic detection), bulk and surface charge screening,~\cite{Broadway_NatElec_2018,Stacey_AMI_2019} and complications from strain interactions, electric-field sensing may prove useful for the chemical sciences. For example, the NV is capable of mapping a single electron \textit{via} its electric field from a distance of roughly $25$ nm~\cite{Dolde_PRL_2014} and of detecting surface electrons at even smaller distances.~\cite{Kim_PRL_2015,Li_PRL_2020} Consequently, improved surface preparation techniques may enable sensing of electrons involved in various chemical processes. Additionally, measurements of local strain~\cite{Marchall_PRAppl_2022} and local band bending~\cite{Broadway_NatElec_2018} would inform the fabrication of stable and shallow NV centers. Furthermore, electric-field sensing could complement magnetic sensing of surface molecules, providing a route to disentangle signal contributions from magnetic fields (arising from electron spins and currents) and electric fields (due to electron charge).

An electric-field sensor can be realized using a single energy-level transition (Sec.~\ref{sec:Magsens}) with a sensitivity defined similarly to those of magnetic-field protocols except that the electric susceptibility parameters ($d_z$ and $d_\perp$) take the place of the gyromagnetic ratio ($\gamma_\mathrm{NV}$). With established techniques for sensing DC and AC electric fields with deeply implanted NV centers,~\cite{Dolde_naturePhys_2011,Iwasaki_ACSNano_2017,Michl_NanoLett_2019} current research efforts are focused on nanoscale sensing of external electric fields with shallow NV centers.~\cite{Oberg_PRAppl_2020,Bian_NatComms_2021,Barson_NanoLett_2021}

In this section, we will highlight two common sensing schemes for measuring electric fields. Both schemes derive from a second order perturbation theory under the assumption of $\hbar \gamma_\mathrm{NV} B_i \ll h D_{\mathrm{gs}}$ and neglecting nuclear spin coupling.~\cite{Dolde_naturePhys_2011,Doherty_PRB_2012} For simplicity, we also ignore contributions from strain; however, we note that in the high-strain limit ($|\vec{\sigma}| \gg |\vec{E}|$), electric-field sensing is only possible along the strain direction.

The first scheme for electric-field sensing requires a weak magnetic bias field along the $z$-axis of the NV center ($B_0 = B_z$), similar to magnetic sensing. In the presence of an electric field, the transition energies are
\begin{equation}
e_{\pm}(B_z, \vec{E}) = e_\pm(B_z) + h d_z E_z(t) \pm \frac{\pi h d_\perp^2}{\gamma_\mathrm{NV}B_z} E_\perp^2(t),
\label{eq:eElec1}
\end{equation}
where $e_\pm(B_z)$ is the contribution from the ZFS and bias field (Eq.~\ref{eq:eMag}) and $E_\perp^2(t) = E_x^2(t) + E_y^2(t)$. By applying sufficiently large $B_z$, it is possible to isolate the $E_z$ component of the field.~\cite{Kim_PRL_2015} 

The second scheme employs a weak perpendicular magnetic bias field ($B_0 = B_\perp = (B_x^2 + B_y^2)^{1/2}$) on the NV center. In this configuration, the transition energies are
\begin{equation}
e_{\pm}(B_\perp, \vec{E}) = e_\pm(B_\perp) + h d_z E_z(t) \mp h d_\perp E_\perp(t)\cos(2 \varphi_B + \varphi_{E}),
\label{eq:eElec2}
\end{equation}
where $e_\pm(B_\perp)$ includes the contribution from the ZFS and a nonlinear shift from $B_\perp$. The additional cosine term involving two in-plane angles ($\tan\varphi_B = B_y/B_x$ and $\tan\varphi_E = E_y(t)/E_x(t)$) allows for control over the electric-field detection axis through variation of $\varphi_B$.~\cite{Dolde_naturePhys_2011} Since $d_\perp/d_z \sim 47$, this detection scheme is significantly more sensitive to perpendicular electric fields; consequently, the $E_z$ term is often excluded for simplicity.

\subsection{Additional sensing modalities}

The NV center is susceptible to additional physical quantities through the dependencies of the Hamiltonian coupling terms; for example, the ZFS splitting is both a function of temperature and pressure ($D_\mathrm{gs}(T, P)$).~\cite{Acosta_PRL_2010,Doherty_PRL_2014} Consequently, careful attention must be taken to correctly attribute energy-level shifts that occur under non-ambient experimental conditions, {\it e.g.,} elevated temperatures or pressures required for specific reactions or temperature changes resulting from exothermic and endothermic reactions. 

Lastly, it is also possible to utilize the excited-state coupling terms, which can greatly differ from those found in the ground-state Hamiltonian.~\cite{Maze_NJP_2011,Doherty_PhysRep_2013} Of particular note, the excited-state electric susceptibility parameters are much larger than those of the ground state, enabling significantly more sensitive electric-field detection.~\cite{Block_PRAppl_2021} However, such techniques employ resonant optical pumping that is only possible at cryogenic temperatures, precluding general use.~\cite{Block_PRAppl_2021}

\subsection{Experimental Considerations}
\label{sec:ExpCon}
As an illustrative example, we explore the experimental considerations for the detection of a single nuclear spin external to the diamond using variance detection. The magnitude of the magnetic variance generated by the Larmor precession of a single nuclear spin at distance $d$ from the NV is~\cite{Lovchinsky_Science_2016}
\begin{align}
B_\mathrm{n}^2=\frac{(\mu_0 \hbar\gamma_\mathrm{n})^2}{16 \pi^2 d^6},
\end{align}
where $\mu_0$ is the vacuum permeability and $\gamma_\mathrm{n}$ is the gyromagnetic ratio of the nuclear spin. Following the calculations in Sec.~\ref{sec:Magsens}, the minimum detectable variance after $N$ repetitions of the spin-echo experiment is
\begin{align}
\delta B^2 = \frac{\pi^2}{\gamma_\mathrm{NV}^2 T_2^2 F \sqrt{N}},
\end{align}
and the expected number of detectable nuclear spins is
\begin{align}
\label{eq:nnuc}
N_\mathrm{\mathrm{nuc}}=\frac{\delta B^2}{B_n^2}=\left(\frac{4\pi^2}{\gamma_\mathrm{NV} \gamma_\mathrm{n} \hbar \mu_0}\right)^2\frac{d^6}{T_2^2 F \sqrt{N}}.
\end{align}
The contributions of parameters $d$, $T_2$, $F$, and $N$ allow us to identify potential avenues for minimizing $N_\mathrm{nuc}$, ideally approaching the single-spin regime of $N_\mathrm{nuc}=1$. 

Eq.~\ref{eq:nnuc} scales with the sixth power of $d$, emphasizing the crucial importance of minimizing the NV--target distance. While the exact dependence of the magnetic-field strength on $d$ varies for different target geometries ($B_\mathrm{rms}^2 \sim 1/d^6$ for a single spin, $B_\mathrm{rms}^2 \sim 1/d^4$ for a two-dimensional spin layer, and $B_\mathrm{rms}^2 \sim 1/d^3$ for a volume of spins),~\cite{Meriles_JCP_2010,Rosskopf_PRL_2014} reducing this distance will always result in a larger field at the NV. 

Eq.~\ref{eq:nnuc} also illustrates the effect of extending $T_2$, which has inspired large bodies of work aimed at fabricating coherent, near-surface emitters (Sec.~\ref{sec:surf}) in addition to development of sensing protocols for extending $T_2$ by mitigating surface noise (Sec.~\ref{sec:MeasStrat}).

Finally, improving the readout fidelity is another avenue for experimental optimization. For off-resonant excitation, the readout fidelity is limited by short readout times ($T_\mathrm{R}\approx400$ ns) due to the transient nature of the spin contrast (Fig. \ref{fig:NV}e), the low radiative emission rate of the NV ($\gamma=2\pi\times13$ MHz), and poor optical collection efficiency ($\xi\sim0.01$ is typical for confocal measurements of bulk diamond). Such schemes are further limited for $C<1$, resulting in an overall readout fidelity of only $F \sim 0.03$ (Eqs.~\ref{eq:F_1} and~~\ref{eq:F_2}). For such a low value, more than $500$ experimental repetitions would be required to achieve $\text{SNR}=1$. Encouragingly, several routes have been explored to increase off-resonant readout fidelity, including nuclear-spin-assisted techniques (obtaining $F=0.5$)~\cite{Lovchinsky_Science_2016} and spin-to-charge conversion~\cite{Shields_PRL_2015} with near-unity fidelity ($F>0.99$).\cite{Hopper_PRB_2016} However, such protocols come at the expense of additional experimental overhead and significantly longer readout times.

The best combination of NV preparation and readout fidelities have been achieved using resonant excitation of spin-dependent optical transitions within the ZPL.~\cite{Robledo_Nature_2011} Below $10$ K, the ZPL transitions of such emitters can be spectrally resolved and optically cycled multiple times before spin flips occur. This cyclicity allows for spin-state readout \textit{via} excitation and the subsequent presence or absence of corresponding emission, yielding $F = 0.97$.~\cite{Hensen_Nature_2015,Humphreys_Nature_2018} Meanwhile, a small degree of spin mixing within the excited state is used as a resource for spin preparation \textit{via} resonant optical pumping, yielding near-perfect preparation efficiencies ($99.8\%$ into $m_s=0$).~\cite{Hensen_Nature_2015} These techniques have been combined to demonstrate imaging of a $27$-spin $^{13}$C cluster within a diamond substrate at cryogenic temperatures.~\cite{Abobeih_Nature_2019} Finally, access to resonant optical transitions can facilitate all-optical spin manipulation,~\cite{Yale_PNAS_2013} albeit with substantially lower fidelity than microwave techniques due to spontaneous emission. Unfortunately, this set of techniques have largely relied on the exceptional optical stability of naturally occurring NV centers located microns below the diamond surface.~\cite{Ruf_NL_2019} Indeed, resonant excitation of near-surface NV centers is not possible due to instability of the ZPL frequency over time referred to as spectral diffusion.~\cite{Santori_Nano_2010,Robledo_PRL_2010} The asymmetry of the nitrogen and vacancy sites (Fig.~\ref{fig:NV}a) leads to inequivalent electric dipole moments in the ground and excited states, which makes it extremely sensitive to electric-field noise on nearby surfaces. Such noise can cause spectral diffusion of the optical transition of many GHz, precluding spectral selectivity within the ZPL fine structure. Consequently, near-surface NV centers required for studying external targets rely on off-resonant excitation.

We conclude by noting that chemical sensing at diamond surfaces may be performed using either single NV centers or ensembles of shallow defects. Single emitters boast nanoscale spatial resolution but are limited in signal by their radiative emission rate $\gamma$ and optical collection efficiency $\xi$.~\cite{Taylor_NaturePhys_2008,Acosta_PRB_2009}
A reduction in excited-state lifetime can be achieved \textit{via} coupling to an optical resonator structure,~\cite{Janitz_Optica_2020} while
higher optical collection efficiencies can be obtained using nanophotonic structures such as waveguides, lenses, cavities, or gratings.\cite{Hausmann_DRM_2010,Hadden_APL_2010,Robledo_Nature_2011,Riedel_PRA_2014,Neu_APL_2014,Momenzadeh_NL_2015,Li_NanoLett_2015,Wan_NL_2018,Riedel_PRX_2017,Jeon_ACSPhotonics_2020,Mccloskey_ACSAMI_2020} 

In contrast to single NV sensors, dense ensembles offers accelerated measurement times due to an increase in optical emission that scales with $N_\mathrm{NV}$, the number of NVs.\cite{Degen_RevModPhys_2017} Moreover, the sensitivity scales with $\sqrt{N_\mathrm{NV}}$ per unit time, with a best predicted sensitivity of \textit{ca}. 250 aT$/(\sqrt{\mathrm{Hz}}\ \mathrm{cm}^{-3/2})$ in the high-density limit.\cite{Taylor_NaturePhys_2008} Higher density, however, may be accompanied by reduced coherence times that ultimately limit sensor utility.\cite{Choi_PRL_2017,Acosta_PRB_2009} Furthermore, the efficient collection of fluorescence from ensembles of NV centers remains difficult; some nanofabricated structures that offer improvement for single NVs may not be suitable for ensembles. Still, alternative strategies that rely on absorptive and dispersive schemes could improve light collection from ensembles.\cite{LeSage_PRB_2012,Jensen_PRL_2014,Clevenson_NatPhys_2015,Wolf_PRX_2015}  Finally, we note that nanostructured diamond surfaces hosting high NV densities benefit from increased sensor--analyte contact area.\cite{Kehayias_NComm_2017}

\section{Surface influence on NV center stability}
\label{sec:surf}
\begin{figure*}
\centering
\includegraphics[height=6.25cm]{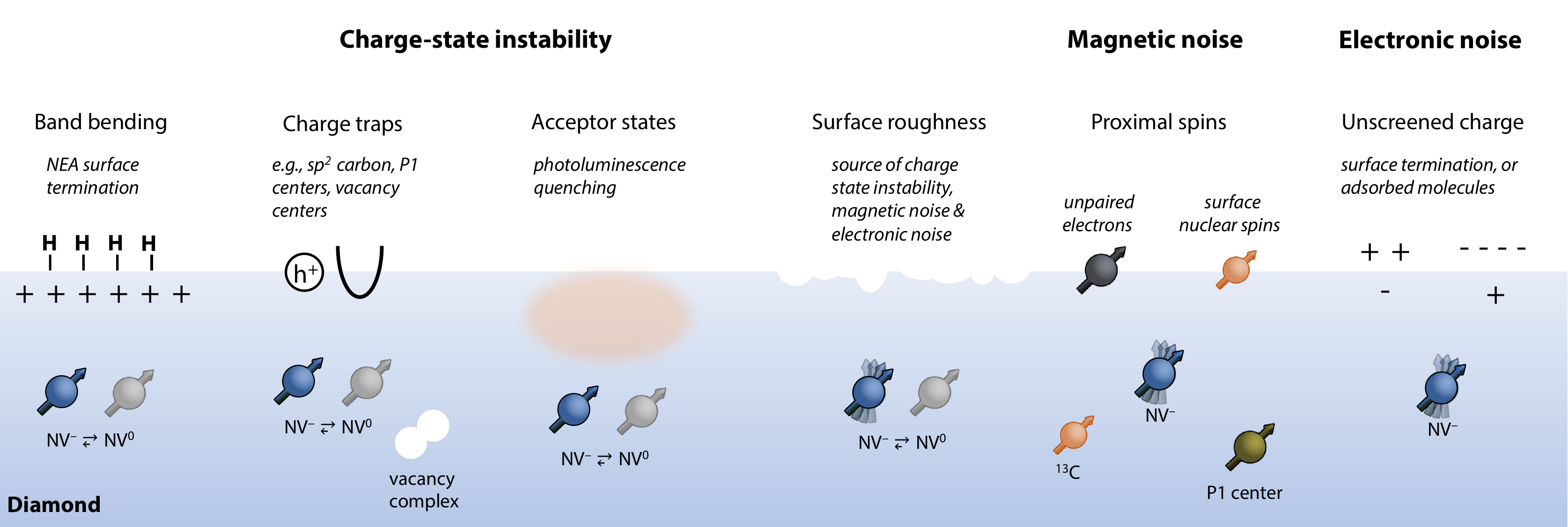}
\caption{Summary of sources of charge-state instability and decoherence for near-surface NV centers.}
\label{fig:instab}
\end{figure*}

Achieving the excellent magnetic-field sensitivity and spatial resolution required for detecting single molecules %
necessitates shallow NV centers (\textit{i.e.}, depths $<$ \textit{ca}. 10 nm). Unfortunately, emitter optical and spin properties deteriorate within \textit{ca.} 100 nm of the diamond surface,\cite{Sangtawesin_PRX_2019} presenting a critical challenge. In this section, we introduce how shallow NV centers are typically generated and then detail the leading sources of their charge instability and spin dephasing (summarized in Fig. \ref{fig:instab}). Finally, we describe suitable surface terminations that have been theoretically and experimentally identified to promote preservation of NV center properties.

\subsection{Generation of shallow NV centers}
\label{sec:nvgen}
Magnetic sensing of external targets necessitates fabrication processes that yield a deterministic density of shallow NVs with high spin coherence.\cite{Smith_Nanophotonics_2019} Most commonly, NV-forming nitrogen atoms are incorporated into electronic-grade diamond through post-growth, blanket implantation,\cite{Rabeau_APL_2006,Pezzagna_IOP_2011,Orwa_JAP_2011} forming a randomly distributed, two-dimensional layer of impurities at a depth and density determined by the acceleration energy and fluence, respectively.\cite{Ziegler_NucInsMeth_2010} In contrast, deterministic impurity placement can be achieved by performing implantation with a focused ion beam (FIB), \cite{Meijer_APL_2005,Lesik_PSA_2013,Tamura_APL_2014,Schroder_NatComm_2017} or through a mask.
\cite{Toyli_NL_2010, Spinicelli_NJP_2011, Bayn_NL_2015,Scarabelli_NL_2016} In fact, a combination of shallow masked implantation and diamond overgrowth has been used to achieve sub-nm placement accuracy for alternative defect centers.\cite{Rugar_NL_2020} 
Subsequent high-temperature, high-vacuum annealing (\textit{ca.} 800 -- 880 \textdegree C) mobilizes vacancies created by implantation, which combine with nitrogen to form NV centers with varying efficiency.\cite{Deak_PRB_2014,Pezzagna_IOP_2011,Naydenov_APL_2010,Chu_NL_2014} Unfortunately, the implantation and annealing process can lead to formation of unwanted vacancy complexes that degrade coherence.
This effect increases with implantation energy, but there is evidence that annealing at higher temperatures (\textit{i.e.,} 1000 -- 1500 \textdegree C) can eliminate localized defects that degrade the spin and optical properties of emitters.\cite{Naydenov_APL_2010,Yamamoto_PRB_2013,Chu_NL_2014,Lang_APL_2020} Furthermore, Fermi-level engineering through co-doping with donor impurities may preclude the formation of vacancy chains.\cite{Luhmann_NatComm_2019} 

An alternative strategy for defect formation is impurity incorporation during diamond growth \textit{via} chemical vapor deposition (CVD) or high-pressure, high-temperature (HPHT) synthesis. In both cases, NV centers can be formed naturally from non-negligible native nitrogen levels. However, such defects occur at random positions within the diamond crystal, limiting their utility. In HPHT synthesis, higher nitrogen content is typically observed (up to hundreds of ppm) than during CVD growth (typically ppb), with varying densities related to the efficiency of nitrogen incorporation along different crystallographic growth directions.\cite{CAPELLI_carbon_2019,Luo_NJP_2022} Conversely, nitrogen incorporation during CVD growth can localize emitters into a single growth layer (referred to as delta doping).\cite{Ohno_APL_2012,Ohno_APL_2014,Osterkamp_APL_2015,Lee_APL_2014,Jaffe_NL_2020} Advantageously, delta-doped emitters may exhibit preferential alignment along one of the four possible crystallographic orientations.\cite{Michl_APL_2014,Ozawa_IOP_2017,TATSUISHI_carbon_2021} Such alignment is useful both for higher experimental throughput in single NV studies as well as for obtaining higher contrast in ensemble NV measurements. Following impurity incorporation, techniques such as electron irradiation, \cite{Martin_APL_1999,Campbell_PSA_2000, McLellan_NL_2016,Ruf_NL_2019} laser writing, \cite{Chen_NatPhot_2017,Sotillo_SciRep_2016} or implantation with additional species\cite{deOliveira_NL_2016,deOliveira_PSSA_2016,Luhmann_NatComm_2019} can be used to generate the vacancies necessary for emitter formation through subsequent annealing. 

\subsection{Origins of instability}
In this section, we explore physical phenomena leading to reduced measurement sensitivity for near-surface NV centers, namely charge instability and magnetic and electronic noise.
\subsubsection{NV charge state conversion.~~}
\label{sec:chargestate}
The NV center has three physically relevant charge configurations comprising the negative ($\mathrm{NV^-}$), neutral ($\mathrm{NV^0}$), and positive ($\mathrm{NV^+}$) states. So far, only the long-lived, $S=1$ ground state (see Sec. \ref{sec:NV}) of the negatively charged state has proven useful for quantum sensing applications. Indeed, the $S=1/2$ ground state of $\mathrm{NV^0}$ is subject to line broadening caused by Jahn-Teller distortion that precludes EPR detection.\cite{Felton_PRB_2008} Moreover, the neutral charge state is also excited by green laser light (ZPL at 575 nm), contributing fluorescent background that reduces measurement contrast. Conversely, the positively charged $\mathrm{NV^+}$ state is expected to be spinless, and is traditionally treated as a dark state since no optical signature has been observed thus far.\cite{Pfender_NL_2017} Proximity to surfaces can exacerbate dynamic or permanent switching to other charge states,\cite{Bluvstein_PRL_2019b}
reducing the charge-state efficiency, defined as
\begin{align}
\zeta=\frac{[\mathrm{NV^-}]}{[\mathrm{NV}^-]+[\mathrm{NV}^0]+[\mathrm{NV}^+]}\approx\frac{[\mathrm{NV^-}]}{[\mathrm{NV}^-]+[\mathrm{NV}^0]}. \label{eq:charge}
\end{align}
While the $\mathrm{NV^0}$ state has garnered interest for super-resolution microscopy,\cite{Han_NL_2010,Waldherr_PRL_2011,Han_NJP_2012,Beha_PRL_2012,Chen__Light2015,Baier_PRL_2020} optical data storage,\cite{Dhomkar_SciAdv_2016,Dhomkar_NL_2018} quantum information processing,\cite{Lozovoi_NatElec_2021} and as an electrically driven single-photon source,\cite{Mizuochi_NatPhot_2012,Doi_PRX_2015} it contributes an undesirable fluorescence background when probing $\mathrm{NV^-}$ centers. Therefore, $\zeta$ should be maximized for magnetic sensing. However, due to unavoidable charge-state cycling between $\mathrm{NV^-}$ and $\mathrm{NV^0}$, the maximum achievable $\zeta$ is limited under normal measurement conditions.\cite{Beha_PRL_2012} 

Photoinduced charge conversion between the $\mathrm{NV^-}$ and $\mathrm{NV^0}$ states can occur \textit{via} one-photon (energy >2.6 eV) or two-photon (energy >1.946 eV) absorption processes that eject an electron into the conduction band of diamond.\cite{Dhomkar_NL_2018} The reverse process, reduction of $\mathrm{NV^0}$ back to its negative state, may also occur \textit{via} both one-photon (energy >2.94 eV) or two-photon (energy >2.156 eV) absorption processes by receiving an electron from the diamond valence band.\cite{Dhomkar_NL_2018} Consequently, green laser excitation that is commonly used for spin preparation and readout results in unavoidable modulation between $\mathrm{NV^-}$ and $\mathrm{NV^0}$, leading to a steady-state $\zeta\approx0.75$.\cite{Waldherr_PRL_2011,Aslam_IOP_2013,Chen_APL_2013} This value can be even smaller for NV ensembles,\cite{MANSON_DiamondRelat_2005} where tunneling between proximal NV centers and nearby nitrogen atoms may occur upon photoexcitation.\cite{Giri_PRB_2018} Multicolor illumination with near-infrared light\cite{Hopper_PRB_2016,Chen_PRApp_2017,Meirzada_PRB_2018} (resulting in spin-dependent charge transfer\cite{Ji_PRB_2016,Roberts_PRB_2019}), doping,\cite{grootBerning_PSSA_2014,Doi_PRB_2016,Murai_APL_2018} and high laser powers\cite{Gorrini_ACSAMI_2021} provide possible routes to increase the relative population of $\mathrm{NV^-}$. Moreover, spin-dependent ionization of NV$^-$ \textit{versus} NV$^0$ states has been used as a resource for achieving single-shot spin readout.\cite{Shields_PRL_2015,Hopper_PRB_2016,Zhang_NatComm_2021} 

Proximal surfaces preferentially convert $\mathrm{NV^-}$ to $\mathrm{NV^0}$ following photoionization since local electronic traps and acceptor states inhibit electron exchange between the NV and the diamond electronic bands (Fig. \ref{fig:instab}).\cite{Bluvstein_PRL_2019b,Chakravarthi_PRB_2021} Moreover, charge-state instability has also been observed in the dark (without photoexcitation), which is hypothesized to occur by tunneling to electron traps.\cite{Dhomkar_NL_2018,Bluvstein_PRL_2019b,Yuan_PRRes_2020} Such non-optical discharge of $\mathrm{NV^-}$ is exacerbated near surfaces, where primal $sp^2$ carbon serves as an efficient charge trap.\cite{Stacey_AMI_2019} 

While it cannot account for all processes leading to NV-center destabilization,\cite{Bradac_NatNanotech_2010,Kaviani_NL_2014,Dhomkar_NL_2018} a band-bending model provides a simplified picture of how the $\mathrm{NV^-}$ state is affected by surface proximity.\cite{Grotz_NatComm_2012} An electric dipole layer at the surface, due to the terminating species or chemical adsorbates, can lead to accumulation of positive or negative charge at the diamond interface. In the case of hole accumulation, electrons are withdrawn from shallow NV centers, yielding the neutral charge state.\cite{Gaebel_APB_2006,Santori_PRB_2009,Fu_APL_2010} This is captured by the electron affinity of the surface, defined as
\begin{align}
\chi=E_{\mathrm{vac}}-E_{\mathrm{CBM}}, \label{eq:eaffin}
\end{align}
where $E_{\mathrm{vac}}$ and $E_{\mathrm{CBM}}$ are the energies of the vacuum level and conduction band minimum, respectively (Table 1). A negative electron affinity (NEA) surface, corresponding to upward bending of the the valence and conduction bands, occurs when $E_{\mathrm{CBM}}$ is shifted above the vacuum level. Such bending is induced by the bond polarization at the surface when adsorbates of lower electronegativity than carbon are present. For instance, hydrogen-terminated diamond is known to exhibit NEA with $\chi=-1.3$ eV measured in vacuum, which depletes the electron density available for near-surface NV centers.\cite{Maier_PRB_2001,Garrido_Langmuir_2008,Hauf_PRB_2011,Newell_JAP_2016} Conversely, a positive electron affinity (PEA) surface occurs when the conduction band minimum lies below $E_{\mathrm{vac}}$, and is generally desirable for stabilizing the $\mathrm{NV^-}$ state. In such cases, chemical terminations or adsorbates with higher electronegativity than carbon increase $\chi$ compared to that of a pristine diamond surface. 
Surface terminations that have been found to achieve PEA for NV sensing are described in detail in Sec. 3.3.

\begin{table*}
\small
\caption{\ Theoretically predicted and experimentally measured electron affinity ($\chi$) values for diamond surfaces with various chemical terminations.}
\label{tbl:example2}
\begin{tabular*}{\textwidth}{@{\extracolsep{\fill}}lllllll}
	\hline
	Orientation & Reconstruction & Termination & Sample preparation & $\chi$ (eV) & Notes  & Ref.\\
	\hline
	(100) & (1$\times$1) & O & theory & +3.91 & ketone termination  & \citenum{Tiwari_PRB_2011}\\
	&   & O & theory & +3.64 & ketone termination & \citenum{Robertson_DRM_1998} \\
	&   & O & theory & +2.63 & ether termination & \citenum{Tiwari_PRB_2011} \\
	&   & O & theory & +2.61 & ether termination & \citenum{Robertson_DRM_1998} \\
	&   & O & thermal oxygen annealing & +2.14 & single crystal & \citenum{Sangtawesin_PRX_2019} \\
	&   & O & chemically oxidized & +1.7 & natural, single-crystal type IIb & \citenum{Maier_PRB_2001} \\
	&   & O & chemically oxidized & \textit{ca.} +1.0--1.5 & natural, single-crystal type IIb & \citenum{BAUMANN_surfsci_1998} \\
	&   & O & plasma oxidation & +0.54 & homoepitaxial boron-doped CVD film & \citenum{WANG_DRM_2000} \\
	&   & H & theory & --3.4 & -- & \citenum{Zhang_PRB_1995} \\
	& (2$\times$1) & N & theory & +3.46 & -- & \citenum{Stacey_AMI_2015}\\
	&   & F & theory & +3.00 & --  & \citenum{Kaviani_NL_2014}\\
	&   & F & thermal dissociation of XeF$_2$ & +2.56 & single-crystal, boron-doped CVD & \citenum{Rietwyk_APL_2013}\\
	&   & O & theory & +2.40 & ether termination & \citenum{Kaviani_NL_2014}\\
	&   & F & theory & +2.13 & --  & \citenum{Tiwari_PRB_2011}\\
	&   & C & UHV annealing & +1.3 & natural, single-crystal type IIb & \citenum{DIEDERICH_SurfSci_1998}\\
	&   & C & theory & +0.8 & -- & \citenum{Zhang_PRB_1995}\\
	&   & C & UHV annealing & +0.75 & natural, single-crystal type IIb & \citenum{BAUMANN_surfsci_1998}\\
	&   & C & theory & +0.61 & -- & \citenum{Tiwari_PRB_2011}\\
	&   & C & theory & +0.51 & -- & \citenum{Robertson_DRM_1998}\\
	&   & C & UHV annealing & +0.50 & natural, single-crystal type IIb & \citenum{Maier_PRB_2001}\\
	&   & H & plasma hydrogenation & +0.19 & homoepitaxial, boron-doped CVD film & \citenum{WANG_DRM_2000}\\
	&   & H/F & theory & --0.38 & 50\% F coverage & \citenum{Tiwari_PRB_2011}\\
	&   & H & hydrogenation \textit{via} hot filament & \textit{ca.} --0.4 & natural, single-crystal type IIb & \citenum{Humphreys_APL_1997}\\
	&   & O & theory & --0.6 & hydroxyl termination & \citenum{Kaviani_NL_2014}\\
	&   & H & plasma hydrogenation & \textit{ca.} --0.8 & natural, single-crystal type IIb & \citenum{BANDIS_SurfSci_1996}\\
	&   & H & plasma hydrogenation & $\le$ --1.0 & natural, single-crystal type IIb & \citenum{DIEDERICH_SurfSci_1998}\\
	&   & H & plasma hydrogenation & --1.3 & natural, single-crystal type IIb & \citenum{Maier_PRB_2001}\\
	&   & H & theory & --1.7 & -- & \citenum{Kaviani_NL_2014}\\
	&   & H & theory & --1.96 & -- & \citenum{Tiwari_PRB_2011}\\
	&   & H & theory & --2.05 & -- & \citenum{Robertson_DRM_1998}\\
	&   & O & theory & --2.13 & hydroxl termination & \citenum{Rutter_PRB_1998}\\ 
	&   & H & UHV annealing & \textit{ca.} --2.2 & single crystal & \citenum{vanderWeide_PRB_1994}\\
	&   & H & theory & --2.2 & -- & \citenum{Zhang_PRB_1995}\\
	&  (2$\times$2) & H/O & theory & +0.5 & mixed H/O/OH termination & \citenum{Kaviani_NL_2014}\\
	&   & H/N & theory & +0.32 & 50\% N coverage & \citenum{Stacey_AMI_2015}\\
	&  -- & O & chemically oxidized & +0.92 & single crystal & \citenum{Sangtawesin_PRX_2019}\\
	(111) &  (1$\times$1) & O & theory & +3.75 & -- & \citenum{Shen_Carbon_2021}\\
	&   & O & theory & +3.42 & ketone termination & \citenum{Shen_Carbon_2021}\\
	&   & N & theory & +3.23 & -- & \citenum{Chou_NL_2017}\\
	&   & F & theory & +2.63 & -- & \citenum{Tiwari_PRB_2011}\\
	&   & O & theory & +1.85 & epoxy termination & \citenum{Shen_Carbon_2021}\\
	&   & C & theory & +1.37 & -- & \citenum{Tiwari_PRB_2011}\\
	&   & H & hydrogenation \textit{via} hot filament & $\le$ +0.7 & natural, single-crystal type IIb & \citenum{Bandis_PRB_1995}\\
	&   & H/F & theory & +0.49 & 50\% F coverage & \citenum{Tiwari_PRB_2011}\\
	&   & H & plasma hydrogenation & < 0 & single crystal, first chemically oxidized & \citenum{BAUMANN_surfsci_1998}\\
	&   & H & plasma hydrogenation & $\le$ --0.9 & natural, single-crystal type IIb & \citenum{DIEDERICH_SurfSci_1998}\\
	&   & H & plasma hydrogenation & --1.27 & natural, single-crystal type IIb & \citenum{Cui_PRB_1999}\\
	&   & H & theory & --1.63 & -- & \citenum{Chou_NL_2017}\\
	&   & H & theory & --2.01 & -- & \citenum{Tiwari_PRB_2011}\\
	&   & H & theory & --2.03 & -- & \citenum{Robertson_DRM_1998,Rutter_PRB_1998}\\
	&  (2$\times$1) & F & theory & +2.49 & -- & \citenum{Tiwari_PRB_2011}\\
	&  & C & UHV annealing (1000 K) & +1.5 & natural, single-crystal type IIb & \citenum{DIEDERICH_SurfSci_1998}\\
	&  & H/F & theory & +0.52 & 50\% F coverage & \citenum{Tiwari_PRB_2011}\\
	&  & C & UHV annealing & +0.5 & natural, single-crystal type IIb & \citenum{Bandis_PRB_1995,BAUMANN_surfsci_1998}\\
	&  & C & UHV annealing (1000 K) & +0.38 & natural, single-crystal type IIb & \citenum{Cui_PRB_1999}\\
	&  & C & theory & +0.35 & -- & \citenum{Robertson_DRM_1998,Rutter_PRB_1998}\\
	&  & C & theory & +0.32 & -- & \citenum{Tiwari_PRB_2011}\\
	&  & H & theory & --2.19 & -- & \citenum{Tiwari_PRB_2011}\\
	& -- & C & UHV annealing (1400 K) & +0.8 & single crystal, graphitized surface & \citenum{Cui_PRB_1999}\\
	(110) &  (1$\times$1) & F & theory & +2.38 & -- & \citenum{Tiwari_PRB_2011}\\
	& & C & theory & +0.91 & -- & \citenum{Tiwari_PRB_2011}\\
	& & C & theory & +0.9 & -- & \citenum{Kern_PRB_1997}\\
	& & H/F & theory & +0.52 & 50\% F coverage & \citenum{Tiwari_PRB_2011}\\
	& & H & theory & --2.41 & -- & \citenum{Tiwari_PRB_2011}\\
	&  (2$\times$1) & H & theory & +2.4 & -- & \citenum{Kern_PRB_1997}\\
	(113) &  (2$\times$1) & N & theory & +3.56 & -- & \citenum{Li_Carbon_2019}\\
	& & F & theory & \textit{ca.} +3.3 & -- & \citenum{Li_Carbon_2019}\\
	& & O & theory & +2.18 & ether termination & \citenum{Li_Carbon_2019}\\
	& & O & theory & --0.06 & hydroxl termination & \citenum{Li_Carbon_2019}\\
	& & H & theory & --1.80 & -- & \citenum{Li_Carbon_2019}\\
	
	\hline
\end{tabular*}
\end{table*}

\subsubsection{Magnetic and electronic noise.~~}

As discussed in Sec. \ref{sec:Magsens}, the best detection sensitivities are obtained by maximizing $T_2$, which is hampered by unwanted noise. For near-surface NV centers, this noise comes from magnetic and electronic sources (Fig. \ref{fig:instab}), though decoherence of shallow emitters is primarily ascribed to magnetic noise.\cite{Sangtawesin_PRX_2019} Deleterious spin impurities occurring in the bulk crystal, such as substitutional nitrogen (P1) centers,\cite{Balasubramanian_NL_2019} can be limited through use of ultra-pure diamond substrates; however, the surface is always problematic, regardless of substrate purity. Indeed, surface spins may contribute considerable magnetic noise through spin flips and precession,\cite{Ajisaka_PRB_2016,Chrostoski_PhysicaB_2021} leading to a reduction in NV-center $T_1$ and $T_2$ \textit{via} dipole--dipole coupling. In both bulk materials\cite{OforiOkai_PRB_2012,Ohashi_NL_2013,Myers_PRL_2014,Rosskopf_PRL_2014,Romach_PRL_2015,Zhang_PRB_2017,Lillie_PhysRevMater_2018,Rollo_PRB_2021} and nanodiamonds,\cite{Panich_EUROPHYB_2006,Tisler_ACSNano_2009,Laraoui_NL_2012,Song_AIPA_2014,Peng_JAP_2020} this magnetic noise may arise from surface nuclear spins, unpaired electrons in dangling bonds, surface adatoms, and molecular adsorbates. Moreover, spin-orbit interactions in \textit{sp}$^2$ carbon, as well as potential biradical spin character or magnetization in mixed \textit{sp}$^2$/\textit{sp}$^3$ material, may also contribute.\cite{Stacey_AMI_2019} 

The influence of a surface spin bath on shallow NV centers can be exploited as a useful probe of the local chemical environment,\cite{Tetienne_PRB_2013,McGuinness_IOP_2013} which has been widely used for chemical and biochemical sensing with NV centers both in bulk diamond and nanodiamond systems. In such experiments, relaxometry measurements that do not require microwave irradiation are used to characterize changes in $T_1$ caused by \textit{e.g.}, proximal paramagnetic species,\cite{Pelliccione_PRAppl_2014} redox reactions,\cite{Barton_ACSNano_2020} changes in pH\cite{Fujisaku_ACSNano_2019}, and radical production.\cite{Martinez_ACSSensors_2020,Nie_NL_2022,WU_redoxbiol_2022} Nevertheless, for highly sensitive techniques such as NMR spectroscopy, these sources of magnetic noise are often undesirable. Consequently, the optimization of diamond surface chemistries is a critical step toward realizing the full potential of NV sensing for dilute assemblies of external spins.

The influence of electric-field noise on the NV is thought to be comparatively minor,\cite{Sangtawesin_PRX_2019} unless operating in a regime where the NV EPR frequencies exhibit anticrossings\cite{Jamonneau_PRB_2016} or in an off-axis magnetic field.\cite{Dolde_naturePhys_2011,Doherty_IOP_2014,Barson_NL_2021}  However, in some cases, electric noise has been shown to dominate magnetic noise.
Indeed, Kim \textit{et al.} showed an almost factor-of-5 increase in $T_2$ for shallow NV centers when bare diamond surfaces were coated in dielectric liquid to shield fluctuating surface charges.\cite{Kim_PRL_2015} In contrast, the magnetic noise generated by dark spins was unaffected by the coating.
In addition, Myers \textit{et al.} used complementary double-quantum relaxometry measurements to differentiate between electric and magnetic noise sources, demonstrating that $T_1$ was limited by surface electric-field noise at low magnetic fields.\cite{Myers_PRL_2017} 

Moving forward, systematic investigations using multiple measurement techniques will be essential in optimizing surface preparations for attaching molecules of interest. For (bio)chemical sensing, functionalization with dense, highly charged, and mobile biomolecules contributes additional sources of electric-field noise and may pose a challenge for highly sensitive measurements. One solution would be to operate at large magnetic-field strengths; in addition, dielectric surface layers (\textit{e.g.}, solid- or solution-phase capping material) may mitigate electric-field-driven decoherence of shallow NV centers.\cite{Kim_PRL_2015,Chrostoski_PRApp_2018} Still, the choice of dielectric coating depends critically on the operational frequency range of the sensor\cite{Chrostoski_PRApp_2018} and its compatibility with the analytes of interest. An as-yet unexplored strategy to shield near-surface NVs from fluctuating electric fields could leverage surface dipole control using self-assembled monolayers (SAMs).\cite{Thomas_ACSNano_2015,Kim_NL_2014} More generally, magnetic and electric noise sources from surface traps could be reduced using highly homogeneous chemical surface functionalization, which is detailed in Section 4. 

\subsection{Surface terminations for increasing NV stability}
Common diamond CVD growth employs hydrogen gas, resulting in hydrogen termination of dangling bonds at interfaces.\cite{Seshan_JCP_2013,Nemanich_MRSBULL_2014} This termination induces a highly unfavorable NEA that destabilizes the $\mathrm{NV^-}$ state. Furthermore, H termination is expected to introduce sub-bandgap states resulting in delocalization and loss of NV electrons upon photoexcitation.\cite{Kaviani_NL_2014} While hydrogen termination will naturally degrade when exposed to air, the systematic chemical control of diamond surfaces offers a solution to combat this destabilization of shallow NV centers.~\cite{Kaviani_NL_2014,Chou_MRSComm_2017} %
Below, we summarize methods for tailoring the diamond surface termination and explore the resulting influence on NV-sensing properties. Subsequent chemical functionalization with molecular films of interest is described in Sec. \ref{sec:func}.

\subsubsection{Fluorine surface termination.~~}
Halogen termination of diamond may occur naturally during CVD growth processes that employ halogenated carbon precursors, or by post-growth treatments including plasma exposure,\cite{Popv_DRM_2008,Denisenko_DRM_2010,Widmann_PSSa_2014} electrochemical techniques,\cite{Wang_carbon_2021} atomic beams,\cite{Freedman_JAP_1994} X-ray irradiation,\cite{Smentkowski_Science_1996} and others.\cite{Ando_DRM_1996,Miller_Langmuir_1996,Miller_SS_1999} In particular, termination with fluorine has received considerable attention in the context of NV sensing; as the most electronegative element, fluorine is a promising surface termination for inducing PEA.\cite{Tiwari_PRB_2011} Moreover, the $100\%$ natural abundance of the $I=1/2$, $^{19}$F isotope would enable its proposed use as a platform for nuclear-spin-based quantum simulation,\cite{Cai_NatPhysics_2013} further motivating its exploration. 

\begin{figure}[h!]
\centering
\includegraphics[height=14cm]{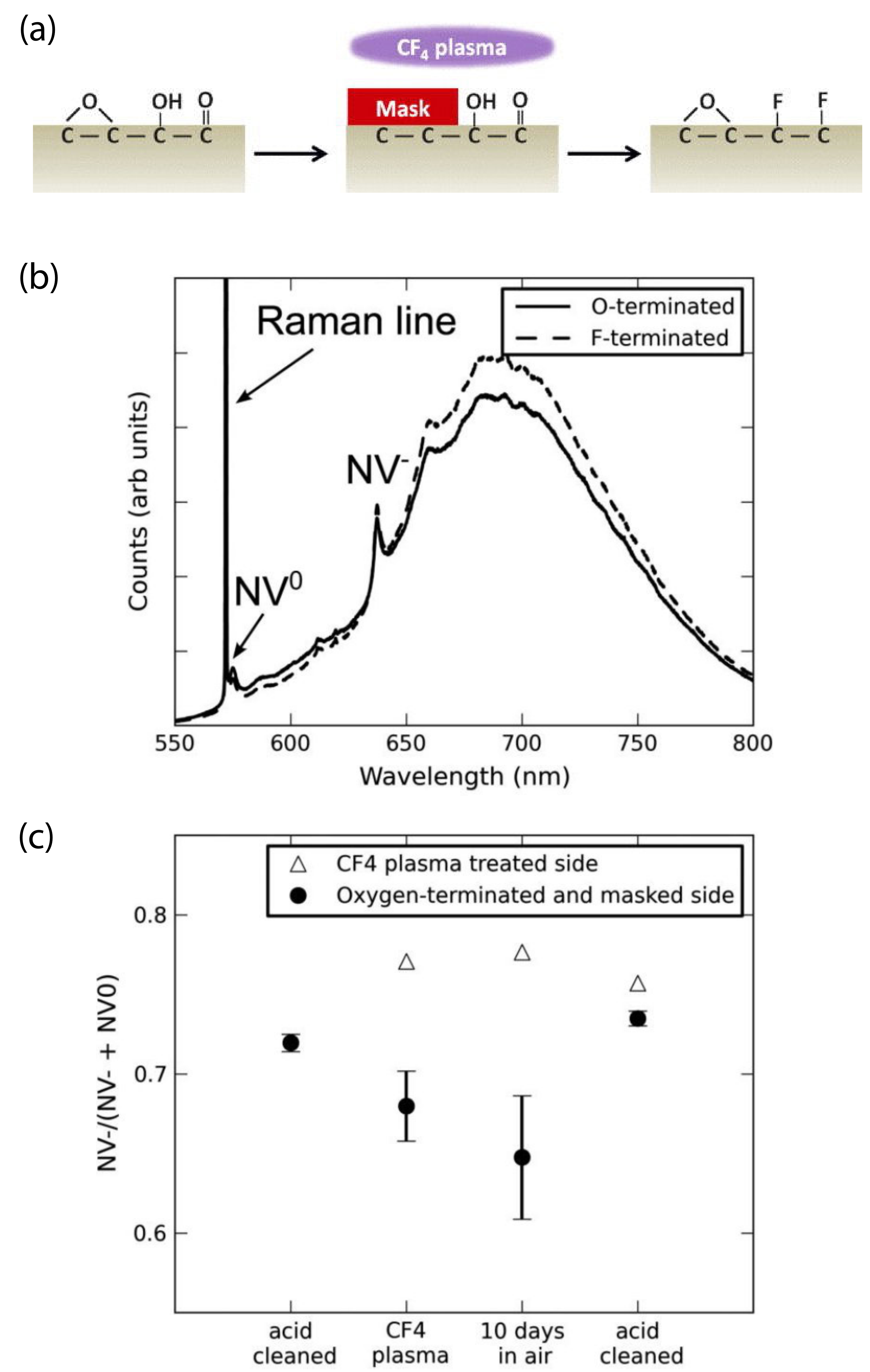}
\caption{Characterization of the impact of fluorine-terminated diamond on shallow NV centers. (a) Schematic showing use of CF$_4$ plasma to induce surface fluorination of oxygen-terminated diamond. (b) Representative photoluminescence spectra comparing oxygen- and fluorine-terminated surfaces. Zero-phonon lines (ZPL) of the NV$^0$ and NV$^-$ are indicated. (c) Ratio of the NV$^-$ ZPL area to that of the sum of NV$^-$ ZPL and NV$^0$ ZPL areas after different surface preparation conditions. Adapted with permission from Ref.~\citenum{Cui_APL_2013}. Copyright 2013 AIP Publishing LLC.}
\label{fig:fluorine}
\end{figure}

Theoretically, fully fluorinated surfaces are expected to yield favorable PEA. For example, (100)-cut diamonds, which are most commonly used, are estimated to have $\chi =2.13-3$ eV.\cite{Tiwari_PRB_2011,Kaviani_NL_2014} Furthermore, (111) surfaces are theoretically predicted to have up to $\chi=2.63$ eV,\cite{Tiwari_PRB_2011} while fluorinated (113) interfaces are estimated to yield $\chi=3.30$ eV.\cite{Li_Carbon_2019} In addition, recent density functional theory calculations suggest that this termination is also highly suitable for the less-studied (110) interface.\cite{SHEN_Carbon_2022}

Experimentally, fluorine termination has been achieved using several techniques. Rietwyk \textit{et al.} utilized exposure to dissociated XeF$_2$, measuring a PEA of $\chi=2.56$ eV for a (100) surface.\cite{Rietwyk_APL_2013} In addition, Cui and Hu tested the influence of fluorine-terminated (100) diamond, prepared using a CF$_4$ plasma (Fig. \ref{fig:fluorine}).\cite{Cui_APL_2013} Analysis with X-ray photoelectron spectroscopy revealed that a 5-minute exposure led to a \textit{ca.} 3-nm-thick polymerized fluorocarbon on the surface. Moreover, charge-state-dependent NV fluorescence measurements revealed a higher $\zeta$ for fluorine-terminated diamond than for oxygen and hydrogen termination. Similar results were obtained for nanodiamonds treated with electron-beam-induced fluorination, which allows for highly localized surface modification.\cite{Shanley_APL_2014} Finally, Osterkamp \textit{et al.} demonstrated the stabilization of shallow (\textit{ca.} 5-nm-deep) NVs using SF$_6$ plasma exposure,
enabling NMR signal detection from protons in immersion oil on the diamond surface.\cite{Osterkamp_APL_2015}

In contrast to these results, Ohashi \textit{et al.} found that a brief (15 s) treatment with CF$_4$ plasma led to permanent bleaching of \textit{ca.} $30\%$ of shallow NV centers.\cite{Ohashi_NL_2013} This experiment underlines the critical importance of reducing acceptor states and charge traps at the diamond surface, even for interfaces with large global PEA.

\subsubsection{Oxygen surface termination.~~}

Oxygen-terminated surfaces offer a highly promising route to stabilize shallow NV centers while allowing for subsequent chemical functionalization. Importantly, we note that coverage by different oxygen-containing functional groups, such as hydroxyl (C--OH), carboxylic acid (COOH), carbonyl (C=O) and ether \mbox{(C--O--C)} groups, impact both $\chi$ and as the presence of deleterious sub-bandgap surface states. Such states may perturb the excited-state energy levels of nearby NV centers, impacting their charge stability.\cite{Kaviani_NL_2014} In addition, we note that different diamond surface planes may also strongly impact $\chi$ and the presence of surface states. Indeed, recent theoretical work by Li \textit{et al.} suggests that the oxygenated (113) diamond surfaces may increase optical stability of near-surface NV centers compared to the commonly used (100)- and (111)-cut diamonds.\cite{Li_Carbon_2019} %

\begin{figure*}
\centering
\includegraphics[height=21.5cm]{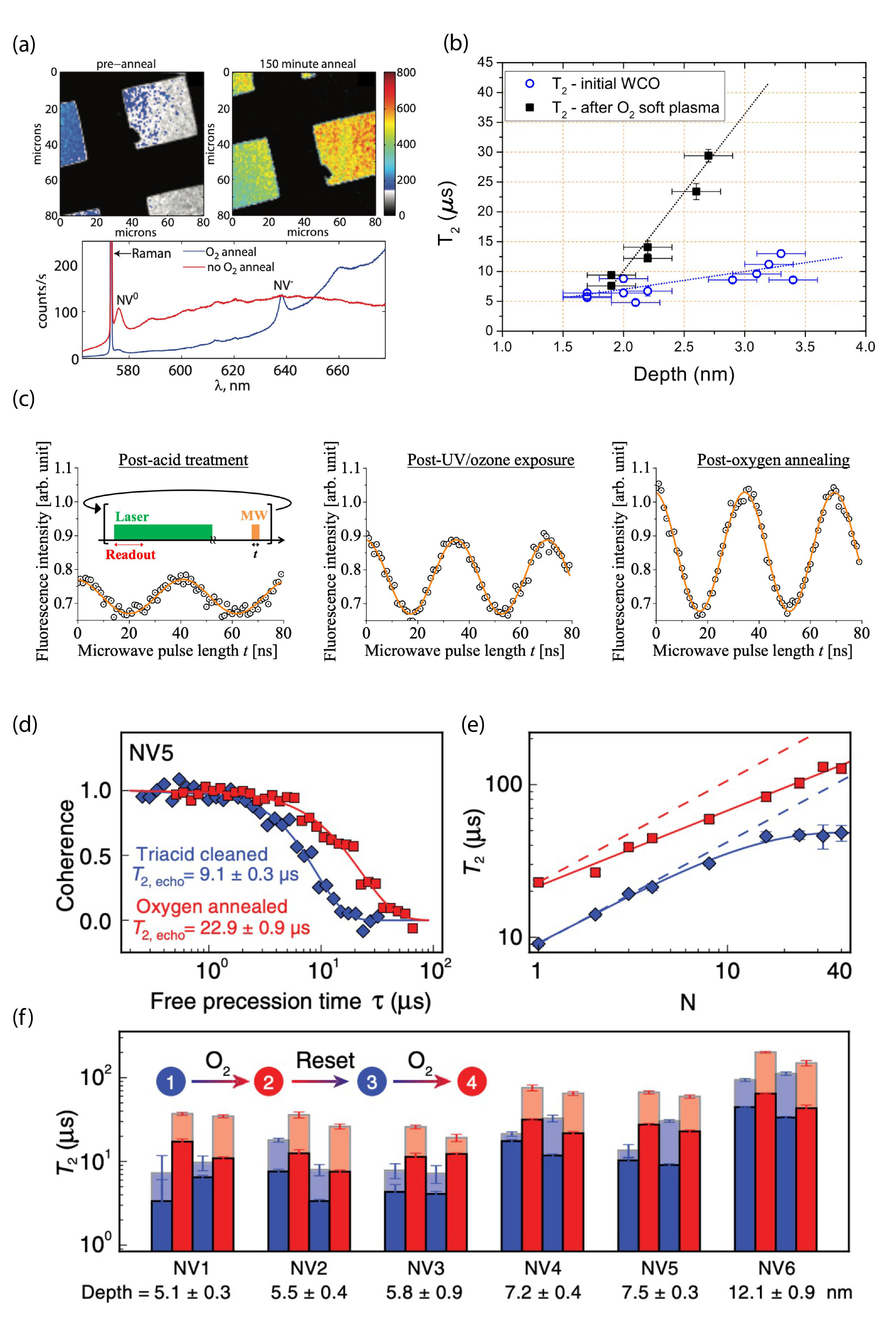}
\caption{Experimental characterization of oxygen-terminated diamond for stabilizing shallow NV centers. (a) Confocal images (top) and photoluminescence spectra (bottom) before and after thermal annealing of diamond surfaces containing shallow NV centers in O$_2$. (b) Spin coherence times $T_2$ (Hahn echo) as a function of depth for NV centers before and after oxygen plasma exposure. (c) Representative Rabi oscillations for a single NV after acid treatment, after UV/ozone treatment, and following thermal annealing in O$_2$. (d) $T_2$ (Hahn echo) decay curves from the same NV center and (e) coherence time as a function of the number of dynamical decoupling pulses before (bue) and after (red) O$_2$ thermal annealing. (f) Reversibility and reproducibility of O$_2$ thermal annealing shown by comparing $T_2$ from Hahn echo (opaque) and XY-8 (transparent). (a) Adapted with permission from Ref.~\citenum{Fu_APL_2010}. Copyright 2010 American Institute of Physics. (b) Adapted with permission from Ref.~\citenum{deOliveira_APL_2015}. Copyright 2015 AIP Publishing LLC. (c) Adapted with permission from Ref.~\citenum{Yamano_IOP_2017}. Copyright 2017 The Japan Society of Applied Physics. (d)--(f) Adapted with permission from Ref.~\citenum{Sangtawesin_PRX_2019}. Copyright 2019 American Physical Society.}
\label{fig:oxygen}
\end{figure*} 

Full termination with hydroxyl groups is theoretically predicted to yield NEA of $\chi=-0.6$ eV and PEA of $\chi=0.2$ eV for (100) and (111) diamond surfaces, respectively,\cite{Chou_MRSComm_2017}
and may yield sub-bandgap states.\cite{Kaviani_NL_2014} Similarly, carboxyl groups can introduce deep, localized acceptor states that quench NV luminescence.\cite{Kaviani_NL_2014} Moreover, ether-like bridges on oxidized (100) surfaces were theoretically calculated to induce a PEA of $\chi=2.4-2.61$ eV;\cite{Kaviani_NL_2014,Tiwari_PRB_2011,Robertson_DRM_1998} 
however, photo-excited electrons from NV$^-$ may be trapped by such surfaces, resulting in blinking.\cite{Kaviani_NL_2014} Encouragingly, epoxy-oxidized (111) surfaces were recently predicted to have PEA of $\chi=1.85$ eV without generation of sub-bandgap states.\cite{Shen_Carbon_2021}
Moving forward, it is likely that a combination of these terminators will yield the optimal combination of PEA and reduced surface states.\cite{Kaviani_NL_2014,Lu_Carbon_2020} 

In practice, oxygen termination results from the use of strongly oxidizing acids (\textit{e.g.}, piranha solution and "tri-acid" mixtures of 1:1:1 H$_{2}$SO$_{4}$:HClO$_{4}$:HNO$_{3}$),\cite{Tisler_ACSNano_2009,Staudacher_Science_2013,Mamin_science_2013,Myers_PRL_2014,Sushkov_PRL_2014,Pham_PRB_2016,Brown_JChemHealthSafe_2019} oxygen plasma,\cite{Hauf_PRB_2011,Kim_APL_2014,deOliveira_APL_2015} thermal annealing in an oxygen atmosphere,\cite{Kim_APL_2014,Yamano_IOP_2017,Sangtawesin_PRX_2019} UV/ozone,\cite{Teraji_JAP_2009} or ozone treatments.\cite{Cataldo_FullNanoCarb_2003} Importantly, such treatments often result in a surface that is decorated by a mixture of oxygen containing functional groups. Various surface characterization methods can be used to distinguish between these terminal groups and to measure their influence on the surface electronic structure\cite{Li_AppSurfSci_2019} including X-ray photoelectron spectroscopy (XPS), ultraviolet photoelectron spectroscopy (UPS), near-edge X-ray absorption fine-structure (NEXAFS) spectroscopy, high-resolution electron-loss spectroscopy (HREELS), and Fourier-transform infrared (FTIR) spectroscopy. However, deterministic control over the relative ratio of these groups is not trivial, and different oxidation procedures yield variable compositions.\cite{Raymakers_JMCC_2019} Moreover, surface roughness and crystallographic orientation also influence the coverage,\cite{WANG_DiamondRelMater_2011,DAMLE_Carbon_2020}. Finally, depending on how harsh the oxidation process is, some procedures may result in undesired graphitization.

Experimentally, surface oxidation treatments have consistently shown improvements in $T_2$ times for near-surface NVs, in addition to higher $\zeta$ both for bulk\cite{Hauf_PRB_2011,Kim_APL_2014,Cui_NL_2015} and nanodiamond materials.\cite{Tsukahara_ACSANM_2019,Ryan_ACSAMI_2018,Rondin_PRB_2010} Fu \textit{et al.} showed efficient conversion of NV$^0$ to NV$^-$ for \textit{ca.} 10 -- 75-nm-deep emitters after annealing at 465 \textdegree C in an oxygen atmosphere (Fig. \ref{fig:oxygen}a).\cite{Fu_APL_2010} Moreover, de Oliveira \textit{et al.} observed a threefold increase in $T_2$ following an O$_2$ soft plasma exposure compared to tri-acid-cleaned surfaces (Fig. \ref{fig:oxygen}b).\cite{deOliveira_APL_2015} In addition, Yamano \textit{et al.} demonstrated an improvement in Rabi oscillation contrast; the $C=0.14$ observed for shallow NV centers in acid-cleaned diamond increased to $C=0.30$ after UV/ozone treatment, further improving to $C=0.43$ following oxygen thermal annealing (Fig. \ref{fig:oxygen}c).\cite{Yamano_IOP_2017} More recently, Sangtawesin \textit{et al.} reported on the impact of thermal oxygen annealing on shallow NV-center spin coherence, along with complementary characterization of the surface morphology (Fig. \ref{fig:oxygen}d-f).\cite{Sangtawesin_PRX_2019} They reported up to a factor-of-four improvement in \textit{T}$_2$ for shallow NV centers compared to samples prepared with tri-acid cleaning alone. These improved properties were attributed to the formation of a highly ordered, predominantly ether-terminated surface. This ordering was facilitated by fabrication of extremely flat (100) surfaces with rms roughness of $<0.4$ nm. In contrast, Braunbeck \textit{et al.} studied the impact of mechanical polishing and etching techniques on coherence times for shallow emitters and found minimal correlation with surface roughness.\cite{BRAUNBECK_dRM_2018} These findings suggest that the particular method of preparing flat diamond surfaces (\textit{i.e.}, etching, polishing, \textit{etc.}) likely has a strong influence on the chemical termination. 

In some cases, surface oxidation can have minimal impact on $T_2$ while having a significant influence on $T_1$ for shallow emitters. For example, Ohashi \textit{et al.} observed no significant change in $T_2$ after oxidation (compared to H-terminated surfaces after tri-acid cleaning) but observed a larger spread in relaxation times ($T_1\approx0.54-4.2$ ms compared to $T_1\approx2.7-3.8$ ms).\cite{Ohashi_NL_2013} Moreover, Tetienne \textit{et al.} saw that thermal annealing in oxygen for 4 h at 465 \textdegree C had little influence on $T_2$ and no effect on the average photoluminescence rate or Rabi contrast.\cite{Tetienne_PRB_2018} In contrast, such annealing resulted in a two-order-of-magnitude reduction in $T_1$ for some emitters; however, longer $T_1$ times could be recovered after tri-acid cleaning. In summary, the experimental results reported for oxygenated diamond surfaces underline the complexity of the relevant spin and charge dynamics that impact shallow NV centers.

\subsubsection{Nitrogen surface termination.~~}

Nitrogen-terminated surfaces have recently received increased attention as an alternative to the more widely employed oxygen treatments. Crucially, the band bending (and resulting electron affinity) at the surface depends sensitively on the bonding nature of the nitrogen atoms.\cite{Pakornchote_PRB_2020,Gong_DiamondRelMater_2021,Korner_PRB_2022} In addition, termination with nitrogen has additional side benefits, including reduced electron-spin noise as well as the potential to create well-defined arrays of nitrogen nuclear spins ($I=1$ or $I=1/2$ for $^{14}$N and $^{15}$N, respectively) that can be probed with NVs.

\begin{figure*}
\centering
\vspace{-1cm}
\includegraphics[height=23.5cm]{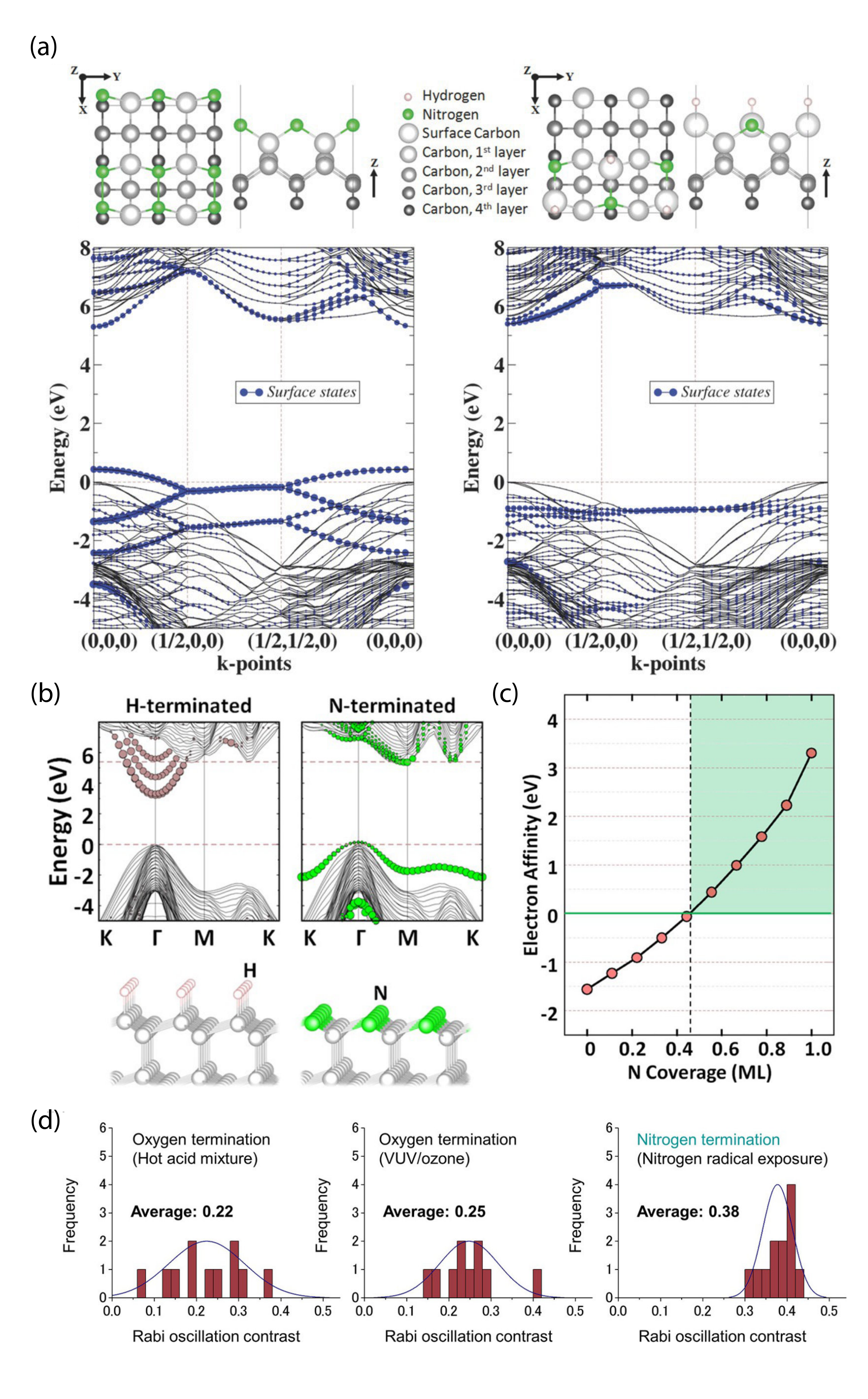}
\vspace{-1cm}
\caption{Theoretical prediction and experimental characterization of nitrogen-terminated diamond for near-surface NV stabilization. (a) Surface termination models and band structure calculations for  fully N-terminated (\textit{left}) and  mixed N/H-terminated reconstructed (\textit{right}) (100) diamond surfaces. Both cases show minimal leakage of surface states into the diamond bandgap. (b) Comparison of H- and N-terminated (111) diamond surfaces. Unlike the N-terminated surface, the H-termination introduces delocalized surface-related image states into the bandgap. (c) Electron affinity as a function of N coverage. (d) Analysis of charge stability for NV centers \textit{via} Rabi oscillations; contrast is compared between oxygen-terminating treatments and after nitrogen termination using nitrogen radical exposure. (a) Adapted with permission from Ref.~\citenum{Stacey_AMI_2015}. Copyright 2015 John Wiley \& Sons, Inc. (b) and (c) Adapted with permission from Ref.~\citenum{Chou_NL_2017}. Copyright 2017 American Chemical Society. (d) Adapted with permission from Ref.~\citenum{Kawai_JPCC_2019}. Copyright 2019 American Chemical Society.}
\label{fig:nitrogen}
\end{figure*}

Theoretical results from Stacey \textit{et al.} showed that $(2\times1)$-reconstructed $(100)$ surfaces that are fully terminated with nitrogen exhibit a PEA of $\chi=3.46$ eV, while $(2\times2)$-reconstructed surfaces with 50/50 N/H termination exhibit a PEA of only $\chi=0.32$ eV (Fig. \ref{fig:nitrogen}a).\cite{Stacey_AMI_2015} Furthermore, Chou \textit{et al.} performed first-principles calculations of $(111)$ diamond surfaces in which terminal carbon atoms could bind to a single hydrogen. With this model, they found that an over $50\%$ substitution of C-H units by isovalent nitrogen led to PEA (Fig. \ref{fig:nitrogen}b,c),\cite{Chou_NL_2017} with full nitrogen coverage resulting in $\chi=3.23$ eV. Interestingly, unlike the (100) and (111) surfaces, nitrogen termination of (113) diamond was proposed as unsuitable for NV sensing since surface-state mixing with the NV excited state would lead to photoionization.\cite{Li_Carbon_2019}

Experimentally, nitrogenation of diamond surfaces has been achieved using anodic polarization in liquid ammonia (requiring conductive, \textit{e.g.}, boron-doped diamond),\cite{Simon_DiamonRelatMater_2009} plasma treatment with nitrogen (N$_2$),\cite{Chandran_APL_2015,Chandran_PSSA_2015,Stacey_AMI_2015,Attrash_JPCC_2020,Attrash_SurfaceScience_2021} ammonia (NH$_3$),\cite{SZUNERITS_ElecComm_2006,Coffinier_Langmuir_2007,ZHU_surfsci_2016,abendroth_arxiv_2022} or mixed sources,\cite{Simon_DiamonRelatMater_2009} radical beam exposure,\cite{Kawai_JPCC_2019} and UV irradiation in NH$_3$ gas.\cite{Zhang_Langmuir_2006}
In these experiments, photoelectron spectroscopies including XPS, NEXAFS, and HREELS were used to probe the binding configuration of nitrogen atoms. The resulting nitrogen may be incorporated as nitrile (CN) groups, imines (C=N), single-bonded (C--N, N--N), or amine (NH$_x$) species. 

In a recent study, Kawai \textit{et al.} demonstrated the stabilization of the $\mathrm{NV^-}$ charge state in shallow defects using nitrogen radical beam exposure (Fig. \ref{fig:nitrogen}d).\cite{Kawai_JPCC_2019} This method is suspected to produce diamond surfaces with a mixture of nitrogen terminations, as characterized by XPS and NEXAFS. Using this technique, the authors reported increased Rabi oscillation contrast compared to NV centers in oxygen-terminated diamond (prepared by hot acid treatment and VUV/ozone exposure). Furthermore, measured $T_2$ values were comparable to prior reports on oxygen-terminated surfaces.\cite{Fukuda_IOP_2018} Subsequently, 4--10 nm-deep emitters were used for NV NMR detection of $^1$H spins in immersion oil.

Interestingly, nitrogenation can also impact the fractional coverage of $sp^2$ and $sp^3$ carbon, a phenomenon that is heavily dependent on the process parameters and diamond surface morphology. For example, an \textit{rf} plasma based on a mixture of NH$_3$/N$_2$ diluted in H$_2$ was shown to induce surface graphitization on polycrystalline, boron-doped diamond.\cite{Simon_DiamonRelatMater_2009} In contrast, we recently showed that exposure of oxygen-terminated, single-crystalline (100) diamond surfaces to NH$_3$ plasma resulted in a reduction of residual $sp^2$ carbon, which may be linked to an observed increase in $T_2$ for shallow NV centers.\cite{abendroth_arxiv_2022} 

Moving forward, derivatization (\textit{e.g.}, of amines) enables complementary tests of chemical functionality. While the introduction of reactive amine functional groups enables facile attachment of molecules, fully amine-terminated diamond surfaces were experimentally shown to exhibit NEA,\cite{ZHU_surfsci_2016} and are thus incompatible with near-surface NV centers. Therefore, mixed surface terminations are likely preferable to maintain NV stability while allowing for covalent molecular attachment.\cite{Stacey_AMI_2015,abendroth_arxiv_2022}\\

\section{Surface functionalization for molecular sensing with NV centers}
\label{sec:func}
NV-center detection of external molecular targets requires both molecular immobilization on the diamond surface and preservation of emitter properties \textit{via} the aforementioned techniques. While the detection of magnetic noise arising from electron \cite{Simpson_NatComm_2017,Wang_SciAdv_2019} and nuclear spins\cite{Staudacher_Science_2013,Mamin_science_2013,Staudacher_Ncomm_2015,Osterkamp_APL_2015,Aslam_Science_2017,Glenn_Nature_2018,Kawai_JPCC_2019,Arunkumar_PRXQuant_2021} external to the diamond has been demonstrated, robust and deterministic functionalization of the diamond surface will prove critical for future experiments. Indeed, such functionalization could facilitate observation of, \textit{e.g.}, surface reactions, conformational changes in biomolecules, spin-dependent effects in charge transfer, and target binding to receptors (such as antibodies or aptamers). Moreover, directed and self-assembly of molecules on surfaces provides well-defined specificity and selectivity, as well as control over molecular orientation (\textit{e.g.}, the availability of binding sites or functional groups) and density, further motivating its development.

Despite being widely regarded as chemically inert, numerous techniques exist to functionalize diamond surfaces, which are the focus of recent reviews.\cite{Szunerits_MRSBULL_2014,Raymakers_JMCC_2019} However, not all strategies for molecular attachment to diamond are compatible with high-quality, shallow NV centers; for example, electrochemical methods, such as diazonium grafting,\cite{Wang_Langmuir_2004,Pinson_CSR_2005} require conductive (\textit{e.g.}, boron-doped) diamond. These substrates are typically incompatible with NV sensing experiments, which require ultra-high-purity substrates. 
In addition, while direct modification of hydrogen-terminated diamond is possible,\cite{Miller_Langmuir_1996,Yang_natmater_2002,Hartl_NatMater_2004,Nichols_JPCB_2005,Colavita_JACS_2007,Stavis_PNAS_2011} it is not be suitable for charge stabilization of near-surface NV$^-$ since local band bending promotes the neutral charge state.\cite{Hauf_PRB_2011} In this section, we explore strategies for stable non-covalent and covalent grafting of molecules to diamond that are compatible with near-surface NV centers. In particular, we focus on covalent attachment for oxygen- and nitrogen-terminated surfaces, in addition to molecular self-assembly on oxide adhesion layers that are grown on diamond.

\subsection{Noncovalent Functionalization} 

One functionalization approach employs non-covalent attachment by physisorption, which is stabilized by, \textit{e.g.}, hydrogen-bond interactions or electrostatic attraction. Alternatively, large molecules of interest (\textit{e.g.}, proteins) could be immobilized within solid polymer matrices that restrict translational and rotational degrees of freedom. Indeed, such a route may prove useful for NV-NMR-based approaches for elucidation of structural information. Moreover, solid supports could take advantage of dielectric shielding to mitigate the impact of electronic noise on NV-center sensors.

\begin{figure*}
\centering
\includegraphics[height=17cm]{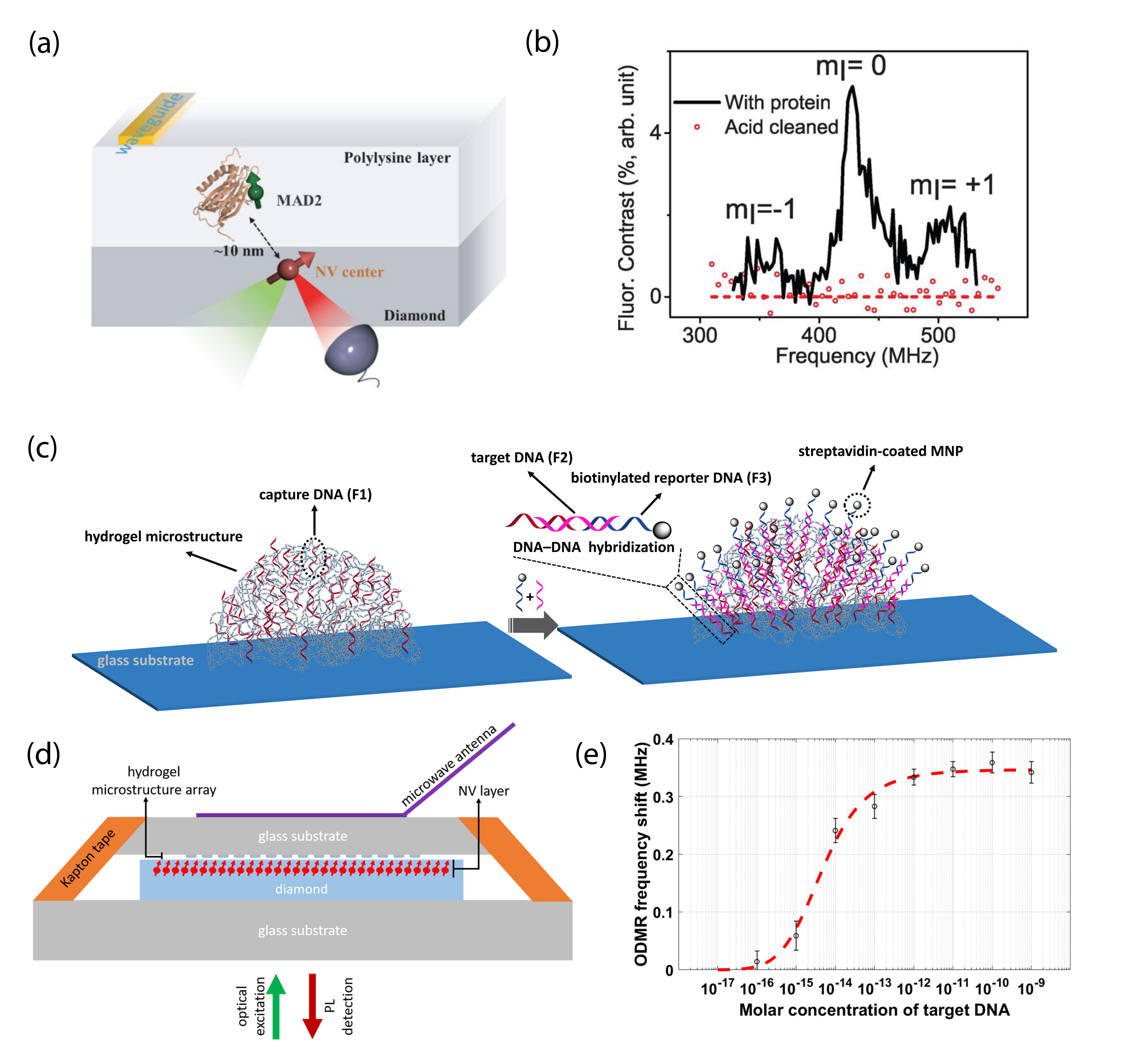}
\caption{NV studies with non-covalently modified surfaces. (a) Schematic of proteins labeled with nitroxide spins and dispersed within a polylysine matrix layer on a diamond surface. (b) EPR spectrum characteristic of the nitroxide spin label tethered to proteins under study. Signals disappear after removal of the proteins from the surface. (c) Illustration of probe DNA immobilization in hydrogel microstructures on glass substrates and hybridization with complementary strands bound to magnetic nanoparticles. (d) Scheme for NV-center detection of magnetic fields arising from magnetic nanoparticles upon DNA capture. (e) Dose response curve of shifts in NV ODMR frequency for increasing concentration of target DNA. (a) and (b) Adapted with permission from Ref.~\citenum{Shi_Science_2015}. Copyright 2015 American Association for the Advancement of Science. (c)-(e) Adapted with permission from Ref.~\citenum{Kayci_PNAS_2021}. Copyright 2021 National Academy of Science.}
\label{fig:noncov}
\end{figure*}

For example, Shi \textit{et al.} used a polylysine matrix to immobilize mitotic arrest deficient-2 (MAD2) proteins that were modified with nitroxide spin labels.\cite{Shi_Science_2015} These molecules were drop-cast onto the diamond surface before freeze drying with liquid nitrogen (Fig. \ref{fig:noncov}a,b).\cite{Shi_Science_2015} Shallow NV centers were subsequently used to monitor dilute densities of these proteins using EPR measurements. Notably, even though motion was restricted by the polylysine matrix, protein dynamics persisted on the millisecond time scale, resulting in spectral broadening of the EPR spectra. Consequently, improved surface functionalization and NV sensing schemes are required for detection of hierarchical ordering and conformational changes in biomolecules. 

Interestingly, Kayci \textit{et al.} recently developed a sandwich bioassay that bypasses diamond functionalization in favor of modifying glass substrates with indexed droplets of hydrogel microstructures. These substrates are then brought into close proximity to a diamond containing near-surface NV centers for sensing (Fig. \ref{fig:noncov}c-e).\cite{Kayci_PNAS_2021} This approach relied on magnetic-nanoparticle-tagged DNA for nucleic acid detection; briefly, poly(ethylene glycol) diacrylate-based hydrogel microstructure networks were polymerized with acrylamide-functionalized oligonucleotides. Complementary target DNA (conjugated with magnetic nanoparticles) then hybridized with probe sequences, resulting in a modified NV ODMR signal. Three-dimensional hydrogel structures hosting bioreceptors could also be patterned directly on planar diamond substrates for applications necessitating direct diamond surface modification. 

Overall, non-covalent functionalization is attractive due to the ease of achieving physisorption compared to chemisorption, particularly for diamond substrates where the density of functional groups may be limited. Nevertheless, under normal sensing conditions, desorption of physisorbed molecules occurs more readily than covalently tethered species. Furthermore, non-specific binding may pose a significant issue when relying on non-covalent attachment methods alone.\cite{Frutiger_ChemRev_2021}

\subsection{Covalent Functionalization}

\begin{figure*}
\centering
\includegraphics[height=6.6cm]{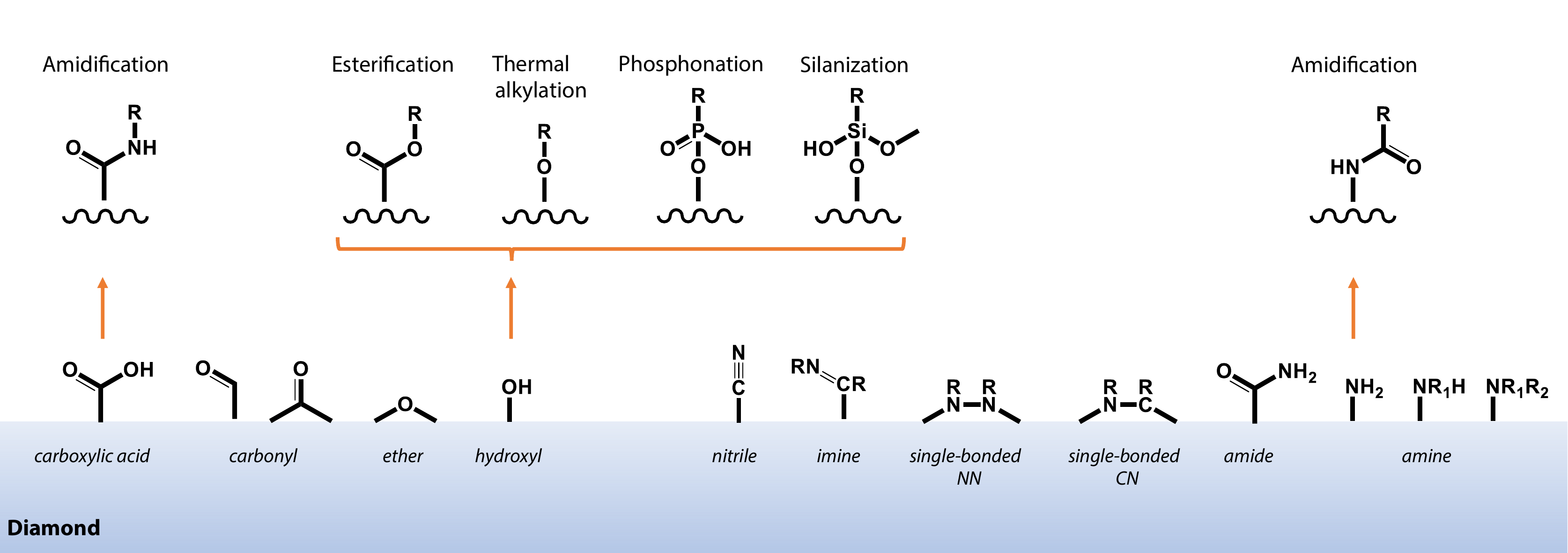}
\caption{Schematic depiction of common functional groups at the surface of oxygen- and nitrogen-terminated diamond, and selected pathways for covalent functionalization as described in detail in the main text.}
\label{fig:O-N-Fun}
\end{figure*}

In contrast to non-covalent techniques, chemisorption offers greater stability and control over molecular attachment to surfaces that exhibit well-defined chemical terminations. 
In practice, the optimal chemical modification procedure and starting diamond surface termination are dependent on the available functional groups of the target molecule. Fig. \ref{fig:O-N-Fun} highlights some selected pathways to functionalize oxygen- and nitrogen-terminated diamond surfaces, which are of great interest due to their compatibility with near-surface NV centers for sensing applications.

\subsubsection{Functionalizing oxygen-terminated diamond.~~}

Conveniently, residual carboxylic acid groups resulting from surface oxidation may be functionalized directly.\cite{Wang_ACSAMI_2012,DAMLE_Carbon_2020}
However, the fraction of carboxylic acid groups is typically low,\cite{Wolcott_JPCC_2014} resulting in dilute attachment that is most suitable for single- to few-molecule experiments. 

\begin{figure*}
\centering
\includegraphics[height=19.7cm]{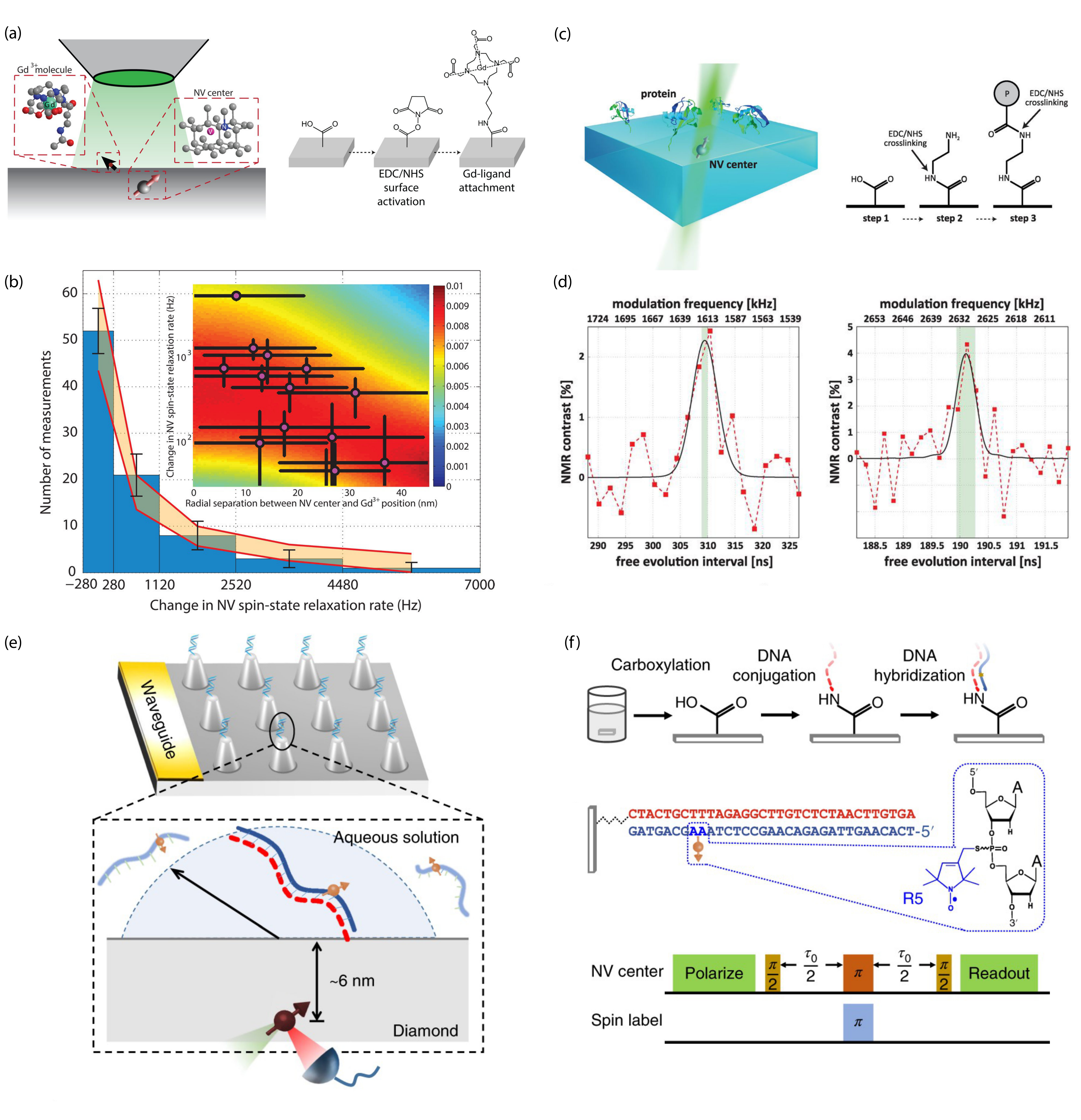}
\caption{Characterization of chemically functionalized, oxygen-terminated diamond surfaces for shallow NV centers. (a) Schematic of single-electron-spin detection from isolated Gd$^{3+}$-chelated molecules bound to carboxylic acids on the diamond surface. (b) Magnetic noise measurements from Gd$^{3+}$-chelated molecules. (c) Schematic for probing isolated proteins by NMR detection with single NV centers (left) and chemical attachment strategy with EDC/NHS crosslinking (right). (d) $^2$H (left) and $^{13}$C (right) detection by NV-NMR of proteins enriched with $^2$H and $^{13}$C isotopes. (e) Schematic of diamond nanopillar arrays functionalized with DNA probes attached to carboxylic acid groups on the diamond surface. (f) Double electron--electron resonance (DEER) pulse sequence used for electron spin detection in hybridized complementary DNA strands. (a) and (b) Adapted with permission from Ref.~\citenum{Sushkov_NL_2014}. Copyright 2014 American Chemical Society. (c) and (d) Adapted with permission from Ref.~\citenum{Lovchinsky_Science_2016}. Copyright 2016 American Association for the Advancement of Science. (e) and (f) Adapted with permission from Ref.~\citenum{Shi_natmethods_2018}. Copyright 2018 Springer Nature Limited.}
\label{fig:oxygenNV}
\end{figure*}

For example, Sushkov \textit{et al.} employed a 1-ethyl-3-(3-dimethylaminopropyl)carbodiimide (EDC)/\textit{N}-hydroxysuccinimide (NHS) crosslinking chemistry to tether paramagnetic molecules composed of a gadolinium ion (Gd$^{3+}$) chelated by an amine-terminated ligand to carboxylic-acid moieties on diamond
(Fig. \ref{fig:oxygenNV}a,b).\cite{Sushkov_NL_2014} Co-localization of isolated molecules using single NV centers allowed for detection of an individual $S=7/2$ electron spin from a (Gd$^{3+}$) chelated molecule using $T_1$ relaxometry. 
Lovchinsky \textit{et al.} later used the same crosslinking chemistry to immobilize ubiquitin proteins \textit{via} exposed carboxylic acid groups on the protein (Fig. \ref{fig:oxygenNV}c,d).\cite{Lovchinsky_Science_2016} This dilute attachment, in combination with enrichment of the protein with $^2$H and $^{13}$C, allowed the authors to detect proteins on the surface using single, shallow NV centers.  
Moving forward, more precise control over specific binding to orient proteins (\textit{e.g.}, antibodies), such as labeling with His-tags, or targeting thiolated cysteine residues, will prove highly useful. 
In particular, specific binding will enable subsequent target attachment to immobilized receptors, as well as positioning of active sites in close proximity to the surface. 

Oligonucleotide probes may also be tethered using readily available modifications to the ends of the sugar phosphate backbone. For example, Shi \textit{et al.} functionalized carboxylic acid groups on diamond nanopillars hosting single NV centers with NH$_2$-modified probe sequences. They subsequently detected hybridization with nitroxide-labeled complementary strands using EPR measurements (Fig. \ref{fig:oxygenNV}e,f).\cite{Shi_natmethods_2018}

Finally, hydroxl groups can be used to attach molecules of interest by esterification\cite{Ushizawa_CPL_2002,Krysinski_JACS_2003,Wang_JPCC_2009,Marcon_Langmuir_2010} and thermal alkylation.\cite{Hoeb_Langmuir_2010} Moreover, hydroxl groups may serve as anchoring points for phosphonic acid headgroups,\cite{Caterino_ACSAMI_2014} for silanization,\cite{Notsu_ElecSSLetters_2001,Ohta_JVacSci_2004,Akiel_JPCB_2016} or for precursor molecules in growth of capping films \textit{via} atomic layer deposition (see Sec. 4.2.3). Starting with oxygen-annealed diamond substrates, Grotz \textit{et al.} used silanization to covalently bind the 4-maleimido-TEMPO spin label.\cite{Grotz_NewJPhys_2011} The authors employed DEER techniques to monitor Rabi nutations of surface-bound electron spins and measured the coupling strength between spin labels and NV-center probes. 
In addition, we demonstrated surface NMR detection using vapor-deposited trimethoxy(3,3,3-trifluoropropyl)silane, as well as (3-aminopropyl)trimethoxysilane, that were subsequently exposed to trifluoromethyl tags for amine-reactive crosslinking.\cite{abendroth_arxiv_2022} Signal from $^{19}$F spins were readily detected \textit{via} NV NMR, however, multilayer film formation could not be ruled out. Generally, preventing multilayer formation or silane film degradation under aqueous conditions or air exposure is difficult and the mechanism for molecular assembly on oxidized surfaces is not fully understood. Still, silanization provides a facile route toward dense surface functionalization.

\subsubsection{Functionalizing nitrogen-terminated diamond.~~}

Nitrogen-terminated diamond surfaces decorated with amine groups can be used to anchor molecules in a variety of ways.\cite{Coffinier_Langmuir_2007} Analogous to crosslinking of molecules hosting amine groups to carboxylic acid moieties on oxygen-terminated diamond, amide bond formation \textit{via} EDC/NHS coupling chemistry can also be used in the reverse way for amine-terminated substrates.\cite{Wang_ACSAMI_2012,Wang_Langmuir_2012} Such experiments use bifunctional linkers to expose COOH groups on initially amine-terminated surfaces for subsequent NHS/EDC reactions.\cite{Yang_Langmuir_2006b} Amine-to-sulfydryl crosslinking with \textit{m}-maleimidobenzoyl-\textit{N}-hydroxysuccinimide ester is then used for directed attachment, \textit{e.g.}, of thiolated oligonucleotide probes and peptides or proteins containing cysteine residues. Indeed, aminolysis with 4-pentynoic acid in the presence of \textit{N},\textit{N}'-dicyclohexylcarbodiimide (DCC) and 4-(dimethylamino)-pyridine (DMAP) was shown to enable catalyst-free thiol-yne coupling on diamond.\cite{Meziane_Analchem_2012} Moreover, amine-functionalized molecules may also be crosslinked to amines on the diamond surface using glutaraldehyde.\cite{Zhang_Langmuir_2006}

Recently, we introduced a strategy for direct chemical functionalization of terminal amine groups using mixed nitrogen- and oxygen-terminated (N/O) surfaces prepared using NH$_3$ plasma following a thermal oxygen anneal.\cite{abendroth_arxiv_2022} Importantly, short plasma exposure times (\textit{ca.} 20 s) resulted in increased coherence times for NV centers $<10$ nm from the surface and significantly reduced background fluorescence from the diamond interface. 
Subsequent chemical functionalization of the mixed N/O-terminated diamond surfaces \textit{via} highly generalizable amine-reactive crosslinking allowed for tunable molecular density and recyclable functionalization. 
Finally, NV-NMR measurements were used to detect surface-bound analytes with trifluoromethyl tags in the few-molecule ($<200$ molecules) regime. While promising, this approach is hampered by a decrease in the number of NV centers that exhibited ODMR signal for long NH$_3$ plasma exposure times. The mechanism of this destabilization is not well understood, but may be linked to a concomitant increase of reactive amine groups that can contribute to NEA surfaces.\cite{ZHU_surfsci_2016} Moving forward, situations requiring a high surface density of analytes crosslinked to surface-bound amine groups may benefit from co-doping with electron donors to aid in stabilizing the NV$^-$ state.\cite{Luhmann_NatComm_2019}

\subsubsection{Atomic layer deposition of oxide adhesion films.}

\begin{figure*}[h]
\centering
\includegraphics[height=15cm]{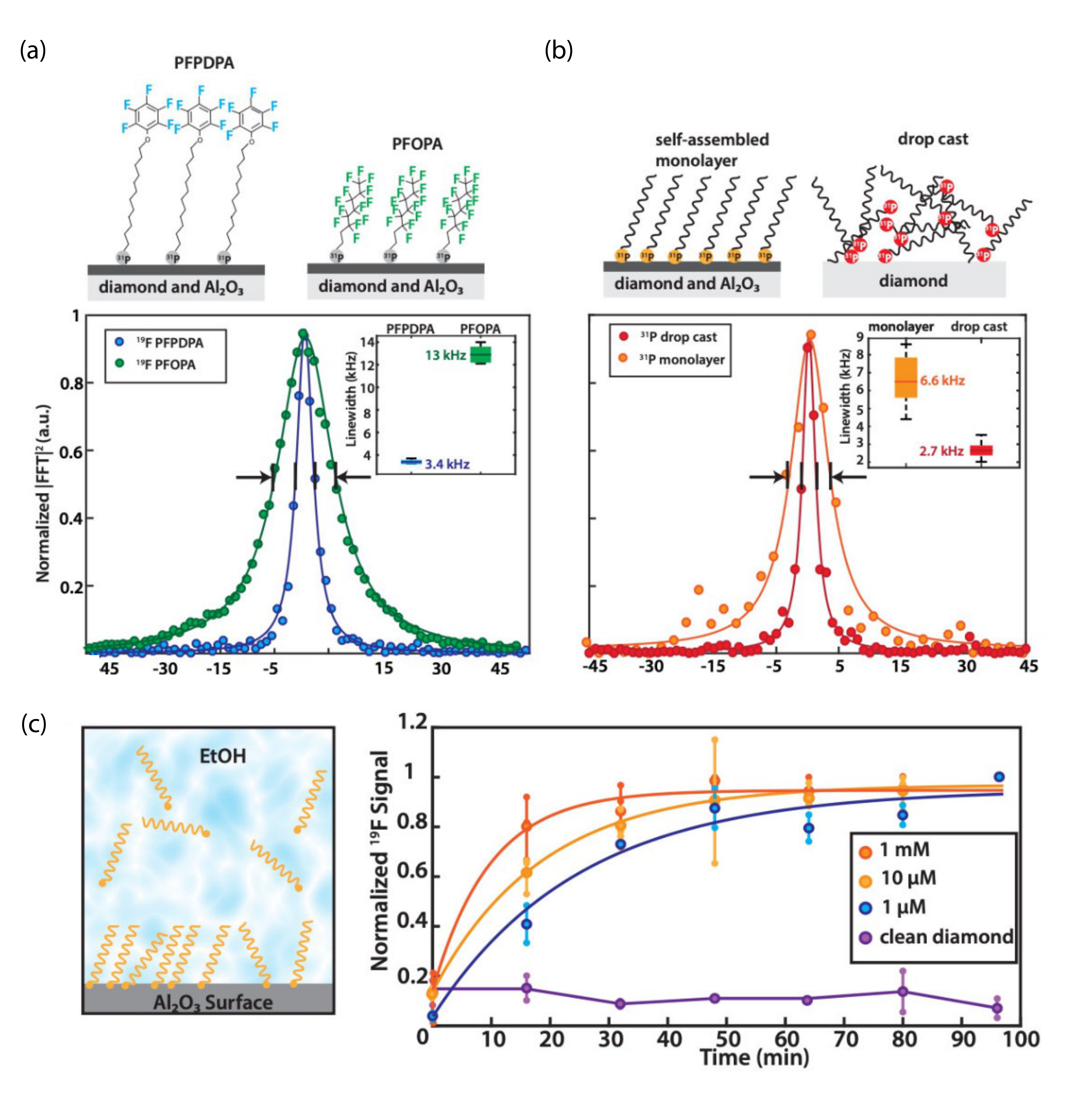}
\vspace{-0.4cm}
\caption{Surface NMR with ensembles of shallow NV centers to probe self-assembled monolayers (SAMs) on diamond capped with Al$_2$O$_3$. (a) Comparison of $^{19}$F resonance linewidth for SAMs of 1H,1H,2H,2H-perfluoroctanephosphonic acid (PFOPA) and 2-pentafluorophenoxydodecylphosphonic acid (PFPDPA). (b) Comparison of $^{31}$P resonance linewidth of SAMs of molecules containing phosphonate headgroups on Al$_2$O$_3$ capping layers \textit{vs} bare diamond surfaces. (c) SAM formation probed by monitoring increase in $^{19}$F resonance signal upon chemical binding of PFPDPA from ethanolic solutions.  Adapted with permission from Ref.~\citenum{Liu_PNAS_2022}. Copyright 2022 National Academy of Science.}
\label{fig:bucher}
\end{figure*}

An alternative to directly binding molecules on the diamond surface employs an intermediate amorphous oxide capping layers that enables straightforward chemical functionalization.\cite{Muller_NatComm_2014} Specifically, atomic layer deposition (ALD) can be used to deposit adhesion films for subsequent molecular attachment \textit{via} well-established oxide chemistries, \textit{e.g.}, with phosphonate or silane headgroups. Recently, Liu \textit{et al.} used ALD to grow \textit{ca.} 1-nm-thick capping layers of Al$_2$O$_3$ on diamond substrates before molecular assembly using phosphonate anchoring groups (Fig. \ref{fig:bucher}).\cite{Liu_PNAS_2022} Introduction of the ALD layer was found to slightly reduce $T_1$ and $T_2$ times for NV centers located \textit{ca.} 5--12 nm below the surface; however, the resulting sensitivity was sufficient to perform ensemble NV-NMR measurements of ${19}$F and $^1$H in molecules during the formation of a self-assembled monolayer. Such real-time observation of molecular signals highlights the exciting potential for NV-NMR spectroscopy to track a wide range of surface chemical reactions over the timescales of minutes to hours.

\begin{figure}[h]
\centering
\includegraphics[height=8cm]{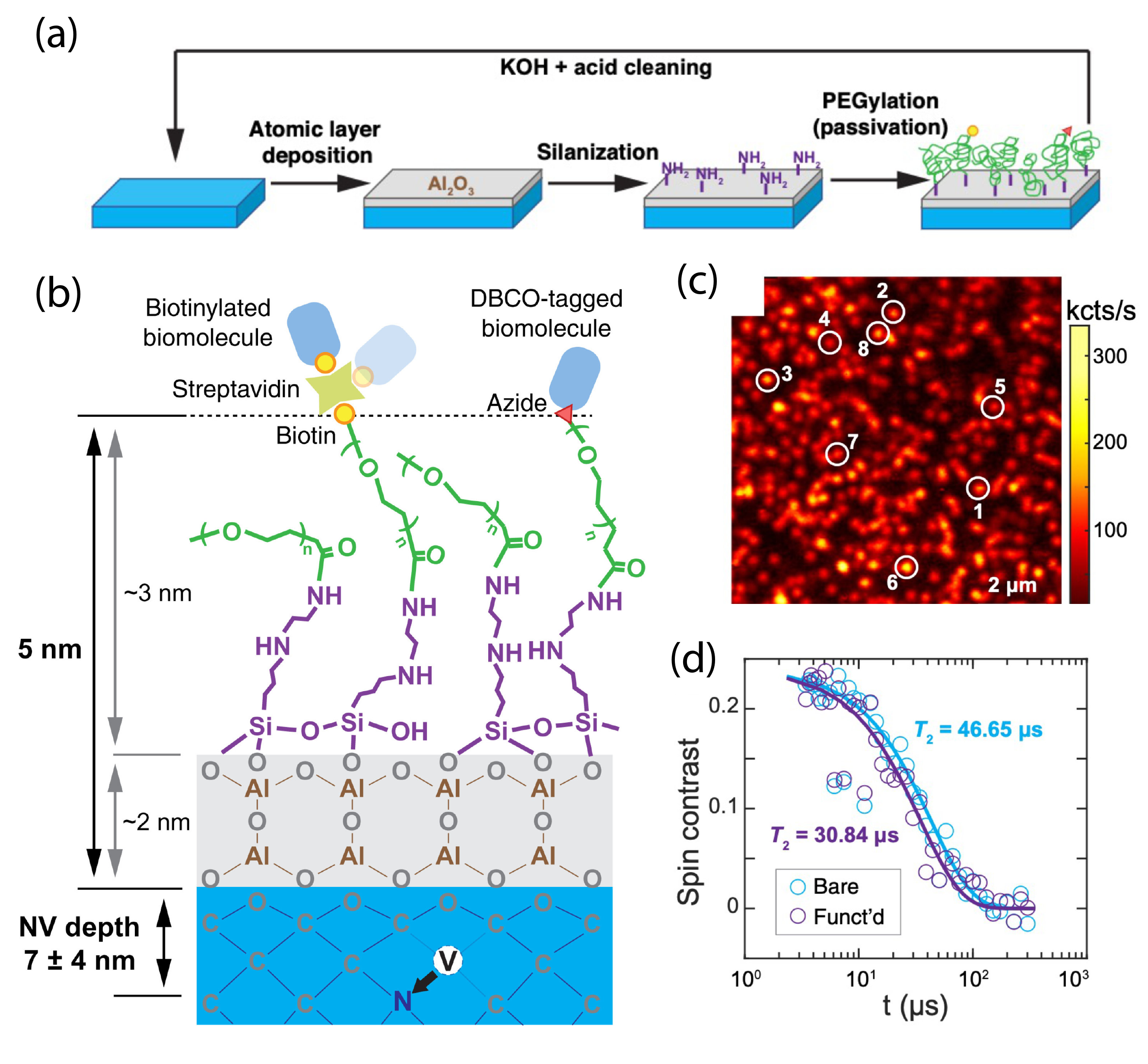}
\caption{Surface functionalization of Al$_2$O$_3$-capped diamond and subsequent characterization of shallow NV properties. (a) Recyclable functionalization scheme showing Al$_2$O$_3$ growth by atomic layer deposition, silanization of the oxide film, and chemical attachment. (b) Illustration of biomolecule attachment on silanized surfaces. (c) Confocal fluorescence image showing near-surface NV centers characterized following surface modification. (d) Representative (YY-8)$_{N=8}$ sensing sequence before and after functionalization for an NV center with depth 4.8 nm. Adapted with permission from Ref.~\citenum{Xie_PNAS_2022} Copyright 2022 National Academy of Science.} 
\label{fig:mauer}
\end{figure}

Similarly, Xie \textit{et al.} used \textit{ca.} 2-nm-thick ALD-grown Al$_2$O$_3$ adhesion layers on oxygen-terminated diamond to test surface functionalization, chemical stability, and preservation of shallow NV sensors (Fig. \ref{fig:mauer}).\cite{Xie_PNAS_2022} Subsequent silanization of these layers enabled grafting of polyethylene glycol (PEG) moieties and attachment of proteins and DNA molecules. Functionalization with PEG molecules is promising as it yields highly biocompatibile surfaces and minimizes non-specific binding.\cite{Lasseter_JACS_2004,Frutiger_ChemRev_2021} While ALD and functionalization resulted in decreased coherence times for shallow (2.3 -- 11 nm deep) NV centers compared to pristine diamond surfaces, some NVs still exhibited coherence times of up to 100 $\mu$s, which are sufficiently long for quantum sensing experiments. Moreover, the authors showed that the Al$_2$O$_3$ layer can be removed by KOH, providing an easily recyclable NV-sensing platform.

In summary, these recent examples show the benefit of amorphous oxide capping layers for NV-center sensing of functionalized diamond surfaces. Specifically, ALD affords precise control of adhesion layer thickness, enables the formation of dense monolayer coverage of molecular adsorbates, and facilitates recyclable surfaces through removal and regrowth. Nevertheless, there are challenges that accompany this functionalization strategy; the adhesion layer leads to increased distance between the NV and target spin, resulting in decreased signal (see Sec. \ref{sec:Magsens}). Moreover, the spin and charge properties of shallow NV centers were observed to degrade with application of ALD layers. Indeed, depending on the stoichiometry of aluminum and oxygen in Al$_x$O$_y$ films, the layers can induce unfavorable electron affinity.\cite{Beattie_DiamondRelatedMaterials_2019} In addition, these films may also host paramagnetic defects leading to reduced $T_2$ times.\cite{Bluhm_PRL_2009,Adelstein_AIPA_2017}

\section{Measurement strategies to mitigate surface noise}
\label{sec:MeasStrat}
In addition to the aforementioned materials optimization and surface chemistries, several strategies can be employed during measurement to mitigate the effects of noise on the NV center. Typically, such measurement sequences increase the sensor coherence time or realize a detection method that is immune to certain noise sources. While a complete overview of the methods for reducing the effects of noise \textit{via} careful engineering of the measurement protocol
(or measurement "hardenings") 
is beyond the scope of this work, we point the reader towards dedicated review articles.~\cite{Barry_RevModPhys_2020,Degen_RevModPhys_2017} In the following section, we discuss illustrative examples for combating unwanted surface effects through experimental implementation. Importantly, the improvements in sensitivity afforded by such techniques are complementary to the advanced surface chemistry discussed thus far. 

\subsection{Dynamical decoupling}
While dynamical decoupling (DD) sequences were not initially developed for sensing, they have proven exceptionally useful for detecting signals at the specific frequencies where they do not refocus the signal~\cite{deLange_Science_2010,Myers_PRL_2014,Du_Nature_2009,Romach_PRL_2015,Degen_RevModPhys_2017}; a phenomenon known as "recoupling". 

The simplest DD sequence is the Hahn-echo protocol~\cite{Hahn_PR_1950}, comprising a single $\pi$ pulse that protects the NV center from dephasing caused by variations in the magnetic field that are slow compared to the experiment duration. The basic idea is illustrated in Figure \ref{fig:sens}: any phase accumulation on the NV center caused by variations in $B_0$ are refocused. Similarly, decoherence from magnetic-field fluctuations at higher frequencies can be mitigated by instead applying a train of $\pi$ pulses, with the interpulse delay $\tau$ dictating the high-frequency cut-off ($1/\tau$) of the noise spectrum. In this case, the magnetic field must remain constant over shorter interpulse delays to achieve a refocusing. As a result, dynamical decoupling protocols can substantially enhance observed coherence times; in contrast, the magnetic noise at $f=1/(2\tau)$ is instead integrated. 

This phenomenon can be exploited for sensing environmental noise \textit{via} changes in coherence at specific frequencies, essentially using the NV as a magnetic spectrum analyzer. Already, such sequences have been used to detect signals originating from AC-modulated single spins,~\cite{Grinolds_NatPhys_2013} currents,~\cite{Pham_NJP_2011,Palm_2022} magnetic gradients~\cite{Hong_NL_2012,Huxter_2022} and electric fields.~\cite{Dolde_naturePhys_2011} For noise with frequency $f$, refocusing will occur for sequences with $\tau=1/(2f)$ as well as for harmonics $\tau'=1/(2N f)$, where $N = 2n + 1$. Moreover, DD sequences may exhibit spurious sub-harmonics for finite-width pulses (dependent on the phase cycling).~\cite{Loretz_PRX_2015} 

Application of over $10^4$ $\pi$ pulses have been successfully applied to a target NV center,~\cite{Zopes_PRL_2017,Abobeih_NatComm_2018} resulting in coherence times exceeding one second at cryogenic temperatures.~\cite{Abobeih_NatComm_2018} It is therefore of crucial importance that DD sequences do not themselves introduce significant errors, necessitating either high-fidelity pulses\cite{Rong_ncomm_2015} or fault-tolerant sequences. A significant reduction in error can be achieved using a scheme proposed by Meiboom and Gill,\cite{Meiboom_RSI_1958} who adapted the Carr-Purcell (CP) sequence 
\begin{align}
\left(\frac{\pi}{2}\right)_x(-\tau/2 - \pi_x - \tau/2-)^n \left(\frac{\pi}{2}\right)_\theta
\end{align}
into a sequence that is now referred to as ``CPMG''
\begin{align}
\left(\frac{\pi}{2}\right)_x(-\tau/2 - \pi_y - \tau/2-)^n \left(\frac{\pi}{2}\right)_\theta,
\end{align}
which provides first-order correction for amplitude, frequency, and timing errors. The phase $\theta$ of the last $\pi/2$ pulse can be tuned to select different sensing schemes~\cite{Degen_RevModPhys_2017} and can prevent 
the accumulation of pulse errors under realistic experimental conditions. Further error correcting protocols of note include the $XYXY$ sequence, the popular $XY-N$ protocols,\cite{Gullion_JMR_1990} and the Knill dynamical decoupling protocol (KDD).~\cite{Souza_RSC_2012}

Such sequences have proven exceptionally useful for detecting, and even controlling, proximal nuclear spins. As with any other source of magnetic noise, DD sequences can very efficiently decouple the NV center
from a bath of proximal spins if they are not also refocused by the DD pulses,~\cite{Abobeih_NatComm_2018} which is typically the case due to the large zero-field splitting of the NV center. Additionally, DD protocols can impart back-action on coupled nuclear spins, which can be used as a resource for realizing coherent control over a multi-spin system.~\cite{Taminiau_Nnano_2014,Bradley_PRX_2019} Encouragingly, recent efforts have made progress in designing DD sequences that refocus noise at only one frequency, suppressing spurious harmonics at the expense of a reduction in signal.~\cite{PhysRevApplied.11.014064, Lazariev2017}
In summary, the detection of nuclear spins \textit{via} DD protocols provides an ideal combination of coherence time enhancement as well as lock-in detection capability for nuclear spins.

\subsection{Coherent driving of surface spins}
Dynamical decoupling alone cannot mitigate noise from residual paramagnetic impurities (\textit{e.g.,} electron spins associated with dangling bonds at the surface or substitutional nitrogen centers) due to their fast Larmor precession at typical magnetic bias fields. However, if such spins have sufficiently long $T_1$ times, one can combat noise by employing continuous, coherent driving of the spin bath.~\cite{deLange_SciRep_2012,Knowles2013} This complementary approach aims at engineering the spectral density of the magnetic noise generated by the electron spin bath itself. In practice, the bath spins are driven with Rabi frequency $\Omega$, which leads to periodic modulation in the resulting noise at the NV location. 
Essentially, such driving shifts the magnetic noise spectrum from $f=0$ to $f=\Omega$, which can reduce the spectral overlap with the NV measurement window,\cite{Degen_RevModPhys_2017} resulting in an increased NV coherence time. Noise can be reduced further by spectral engineering of the applied field\cite{Joos2022} or by using pulsed schemes.\cite{de_Lange_2012,Bauch_PRX_2018} 

Previously, such techniques have been used to reduce dephasing for NV-rich diamonds with large nitrogen content. However, coherent driving of the spin bath is applicable to any paramagnetic spins that exhibit sufficiently long $T_1$ times, including nuclear spins. However, in this case, low nuclear gyromagnetic ratios pose a challenge in achieving high driving frequencies but also reduce the magnetic noise generated by such species. 

\subsection{Repetitive measurement and readout schemes}
Often the sensor coherence time is insufficient for achieving the desired spectral resolution using standard DD protocols; in such cases, weak-measurement techniques may serve as a viable alternative.~\cite{Cujia_Nature_2019,Pfender_NatureComm_2019} Such measurement schemes are appropriate only when coherent detection is possible (\textit{e.g.,} measuring coherently coupled nuclear spins) and involve applying sequential, weak measurements that result in limited back-action on the target spin. Since the NV center can be reinitialized between sequential measurements, the spectral resolution is no longer limited by the NV-center $T_2$ time and instead by that of the target nuclear spin. Importantly, we note that the measurement sensitivity is still limited by the NV $T_2$, motivating the use of highly coherent emitters.

\subsection{Double-quantum sensing}
\label{sec:DQS}
Most measurement protocols leverage only two of the three spin states within the ground-state manifold ($\ket{m_s=0}$ and $\ket{m_s=\pm1}$, see Sec.~\ref{sec:NV}). In contrast, double-quantum protocols make use of the full spin-triplet subspace to obtain higher sensitivity. For example, in the Ramsey-like experiment shown in Fig.~\ref{fig:sens}a, the first $(\pi/2)$ pulse would be replaced by two pulses: a $(\pi/2)_y$ pulse on the $\ket{0}\leftrightarrow\ket{-1}$ transition followed by a $(-\pi)_y$ pulse on the $\ket{0}\leftrightarrow\ket{+1}$ transition. These rotations result in an equal superposition of both the $\ket{-1}$ and $\ket{+1}$ states
\begin{align}
\ket{\psi}=1/\sqrt{2}\left(\ket{-1} + \ket{+1}\right).
\end{align}
During the measurement evolution period, phases are imprinted on both basis states with opposite sign
\begin{equation}
\ket{\psi} = \frac{1}{\sqrt{2}} \left(e^{-\mathrm{i}\gamma_\mathrm{NV}\int B(t)\mathrm{d}t}\ket{-1} +e^{\mathrm{i} \gamma_\mathrm{NV}\int B(t)\mathrm{d}t}\ket{+1} \right)\text{.}
\end{equation}
Finally, this state is mapped back to the starting basis through application of the same pulses in reverse order. The resulting measurement yields double the signal when compared to the single-quantum approaches described in Sec.~\ref{sec:NV}.
An additional benefit of double-quantum techniques are their immunity to common-mode shifts that affect both sub-levels in the same manner. For example, variations of the zero-field-splitting constant caused by temperature shifts,\cite{Acosta_PRL_2010} %
axial electric or strain fields\cite{Bauch_PRX_2018}, and transverse magnetic-field fluctuations.~\cite{Bauch_PRX_2018} If such noise sources dominate the coherence loss, double-quantum DD may yield longer coherence times than single-quantum approaches.
However, for longitudinal magnetic fluctuations, the doubled phase accumulation often comes at the cost of a $\sim 2 \times$ reduction in $T_2$.~\cite{Fang_PRL_2013,Mamin_PRL_2014} 

\section{Emerging Applications} \label{sec:emergingappl}
\begin{figure*}
\centering
\includegraphics[width=0.95\textwidth]{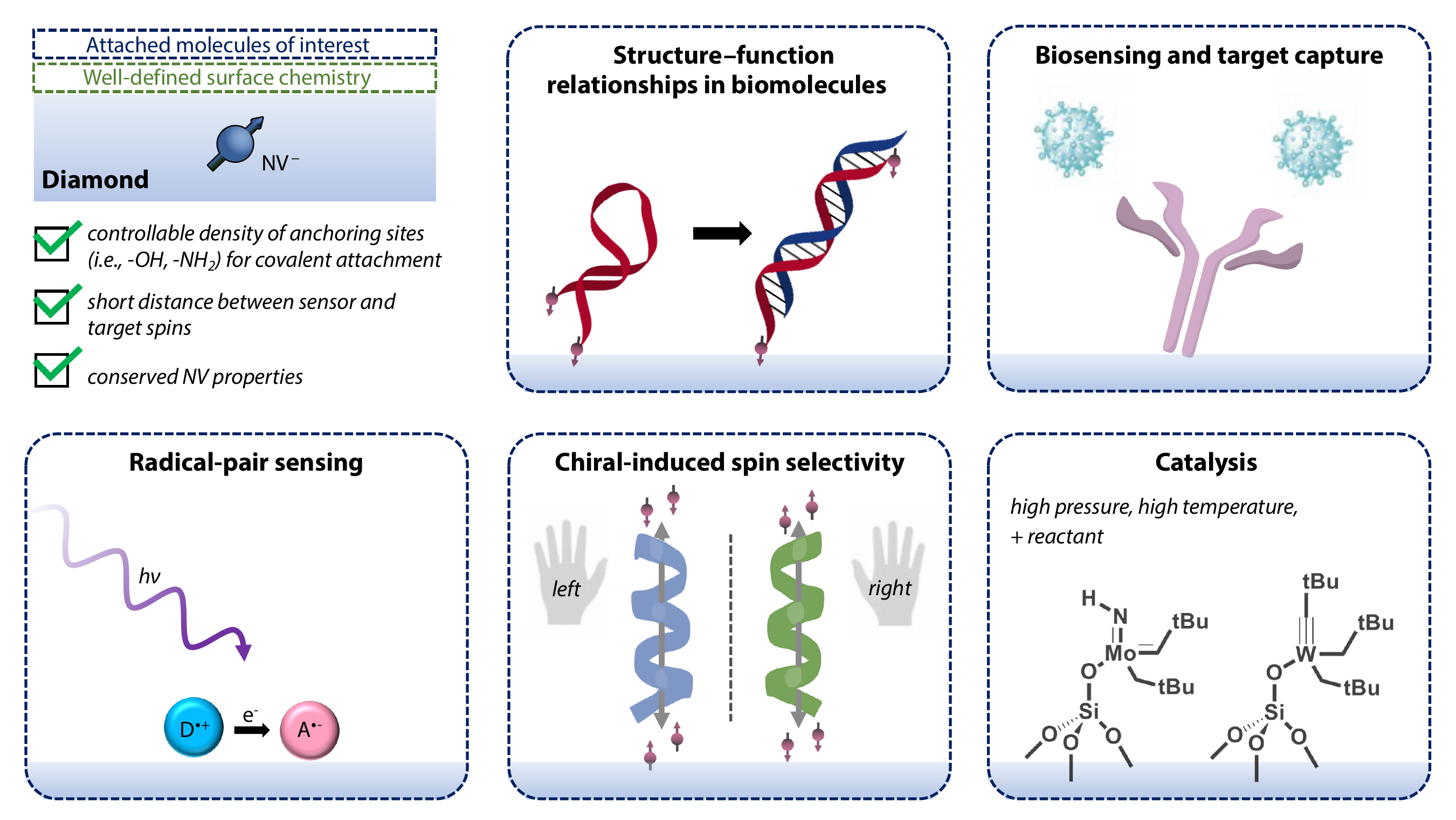}
\caption{Schematic overview of emerging applications for shallow NV centers that are discussed in Sec. \ref{sec:emergingappl}. All  applications highlighted in this Review make use of diamond with well-defined surface termination (\textit{i.e.}, controllable density of anchoring sites and layer thickness as well as conserved NV properties) to which molecules of interest are covalently attached. }
\label{fig:applications}
\end{figure*}

The previously surveyed methods for diamond-surface functionalization could be used in concert with the exceptional sensitivity and spatial resolution of the NV to enable molecular-level detection. 
Indeed, NV sensors could revolutionize many active areas of research since they operate at ambient conditions with minimal perturbation to the system of interest.\cite{Mzyk_AnalChem_2022}
In the following section, we highlight a selection of emerging applications within various fields of chemistry and biology that would benefit from (or necessitate) covalent attachment to a well-defined, functionalized diamond surface (summarized in Fig. \ref{fig:applications}), underscoring the immense potential of the NV-center sensing platform. 

\subsection{Structure--function relationships in biomolecules}

The potential for NV sensors to elucidate the structure of small to hierarchically more complex biomolecules (\textit{e.g.,} proteins) has been widely posited.\cite{Schirhagl_AnnRevPhysChem_2014,Kost_Scirep_2015,Shi_Science_2015,Lovchinsky_Science_2016} While the first demonstration of NV-center NMR of external nuclear spins dates back to 2013,\cite{Mamin_science_2013}\cite{Staudacher_Science_2013} three-dimensional mapping of such spins within a molecule remains an outstanding challenge.

In contrast to conventional solid-state NMR, which has made tremendous progress in structural analysis by employing fast magic-angle spinning (MAS)\cite{Boeckmann_JMagnReson_2015} and high magnetic fields,\cite{Callon_JBiomolNMR_2021} NV NMR has poor spectral resolution. This limited resolution precludes the detection of small chemical shift differences since measurements are typically carried out at low applied magnetic field and MAS cannot (yet) be applied. 
Nevertheless, NV centers could still be used to measure structure-function relationships in biomolecules \textit{via} position-dependent hyperfine interaction with nuclear spins rather than chemical-shift differences.\cite{Zopes_2018,Cujia_NatComms_2022} Furthermore, changes in the magnitude of magnetic noise detected from the Larmor precession of nuclear spin labels can serve as a proxy for conformational changes (\textit{e.g.}, opening of DNA hairpins).

Finally, optical pumping of the NV-center spin can provide a local, low-entropy source of nuclear-spin polarization that could provide complementary structural information to conventional NMR and EPR techniques. In such experiments, nuclear spins coupled to the NV center are polarized beyond the thermal limit by transferring polarization from the NV.\cite{Shagieva_NL_2018,fernandez_NL_2018,Healey_PRA_2021,Tetienne_PRB_2021} Subsequently, the polarized spin bath can be detected classically (using NMR pick-up coils) or coherently using the NV center.\cite{Cujia_Nature_2019}

\subsection{Biosensing}
The biocompatibility of diamond supports the use of NV centers for bio-sensing applications.\cite{Petrini_AdvQuantumTechnol_2020,Zhang_ACSSensors_2021} In addition, there is increased interest in developing biosensors for detection of communicable diseases in light of the COVID-19 pandemic. To that end, the \textit{in vitro} detection of viral RNA using NV centers was first demonstrated in 2020 using a lateral-flow format for HIV diagnosis.\cite{Miller_Nature_2020} In this work, the presence of HIV-1-RNA caused immobilization of antibody functionalized nanodiamonds; this was subsequently observed \textit{via} fluorescent readout of the NVs. In addition, Li \textit{et al.} proposed a sensing scheme for SARS-CoV-2 RNA that exploits $T_1$ relaxation of NV centers in nanodiamonds.\cite{Li_NL_2022} In analogy to earlier work on pH and redox sensing,\cite{Rendler_NatComm_2017} they propose Gd$^{3+}$ as a source for magnetic noise. Specifically, nanodiamonds could be coated with DNA strands containing electrostatically bound Gd$^{3+}$; upon base-pair matching with the complementary viral RNA, the DNA molecules including Gd$^{3+}$ complexes would be released from the diamond surface. This reaction would result in decreased noise and therefore longer relaxation times for near-surface NV centers.

Notably, the majority of NV biosensing research has thus far leveraged nanodiamonds rather than bulk substrates, which can directly utilize many of the diamond surface preparation and functionalization strategies described in this Review. Further supported by significant advancements in production and purification of nanodiamonds, this alternate platform has seen application in biomedical sensing and imaging,\cite{Reineck_ACSNano_2017,McGuinness_NatNano_2011,Kucsko_nature_2013,Ermakova_NL_2013,Nie_SciAdv_2021} drug delivery,\cite{Zhu_Theranostics_2012} and reaction monitoring,\cite{Barton_ACSNano_2020,Bachman_Langmuir_2021} as discussed in recent reviews.\cite{Torelli_Small_2019,Olia_ACSANM_2021,Mzyk_AnalChem_2022} Specifically, these experiments have monitored physiological variables, such as concentration of ions, radicals, and biomolecules, in addition to pH, temperature, and redox potential or forces.\cite{Zhang_ACSSensors_2021}
Informed by these efforts, bulk diamond substrates offer a complementary or alternative sensing platform for hosting NV centers with significantly improved spin properties. It is worth noting that motivations for chemical control of bulk surfaces may differ from those of nanodiamonds intended for studies of aggregation, protein corona formation, cellular uptake, and targeted delivery. In addition, alignment of the magnetic field required for NV-center experiments is simpler for bulk substrates than for randomly orientated nanocrystals.\cite{Holzgrafe_PRA_2020,Igarashi_JACS_2020} Nevertheless, both nano- and bulk-diamond platforms would benefit from improved NV-center properties and deterministic surface functionalization protocols, motivating progress in this field.

\subsection{The radical-pair reaction}
Radical pairs are reaction intermediates that either encounter each other in solution or are created simultaneously, usually by a chemical reaction such as homolytic bond cleavage or photo-induced electron transfer (\textit{e.g.}, in blue-light photoreceptive cryptochromes). \cite{Woodward_PIRKAM_2022}\cite{Biskup_AngewChemIntEd_2009}\cite{Solovyov_ScientificReports_2013}

Research on radical pairs began in the late 20th century in the context of Chemically Induced Dynamic Nuclear Polarization (CIDNP)\cite{Ward_JACS_1967} and Chemically Induced Dynamic Electron Polarization (CIDEP). Later, this research gained interest in the area of magnetic-field effects (MFE),\cite{Steiner_ChemRev_1989}\cite{Buchachenko_AccChemRes_2017} \cite{Steiner_JPhysChemC_2018} including low-field effects with biological significance\cite{Timmel_MolPhys_1998}\cite{ODea_JPhysChemA_2005} (\textit{e.g.}, magnetoreception in birds\cite{Hogben_PhysRevLett_2012}\cite{Hore_AnnuRevBiophys_2016}). Currently, most radical-pair research focuses on spin-correlated states, which may be either a spin singlet or triplet state.\cite{Harvey_JACS_2021} Since radical recombination commonly occurs solely from the singlet state, the inter-conversion between singlet and triplet states, which is mainly driven by differences in Zeeman frequencies and hyperfine interactions with surrounding nuclei, is a crucial parameter for the chemical outcome. 

Recently, Finkler \textit{et al.} proposed to use the NV as a sensor for observing this radical-pair mechanism.\cite{Finkler_PRApp_2021} In such an experiment, a nearby radical pair would experience an effective magnetic field (and therefore singlet-to-triplet conversion rate) that depends on the spin state of the NV sensor. 
An alternate strategy was introduced by Liu \textit{et al.}\cite{Liu_PRL_2017}, who proposed to use the electric-field sensitivity of the NV to measure radical-pair recombination rates. Essentially, the charge-separated state of the pair generates a non-zero electric field at the NV position that vanishes upon recombination. Consequently, detection of this field would enable measurement of recombination rates at the single-molecule level. We note that both proposals rely on radical-pair generation in a covalently attached bio-molecule (\textit{via} photo-induced electron transfer with a blue light source), which would benefit greatly from the advances in diamond functionalization discussed in this Review. 

\subsection{Chiral-induced spin selectivity}

Chiral-induced spin selectivity (CISS) refers to enantioselective and electron spin-dependent transmission of electrons through chiral molecules and crystalline materials.\cite{NaamanWaldeck_AnnuRevBioPhys_2021,Aiello_ACSNano_2022final} While spin-selective electron transfer is typically associated with magnetic materials or those possessing substantial spin-orbit coupling, the discovery of the CISS effect suggests that organic molecules lacking inversion symmetry that are composed of low-atomic-weight building blocks may also be promising systems for spintronics applications.\cite{Ray_Science_1999} Since this discovery, experiments have revealed large asymmetry in the scattering probability of polarized photoelectrons traversing thin organized films of chiral organic molecules,\cite{Ray_Science_1999,Gohler_Science_2011,Kettner_JPCL_2018,Abendroth_JACS_2019,Stemer_NL_2020} observed spin selectivity in the conduction regime,\cite{Abendroth_ACSNano_2017,Mishra_JPCC_2020,Liu_ACSNano_2020,Sang_JPCC_2021} and measured spin-dependent charge polarization within molecules and at surfaces.\cite{Kumar_PNAS_2017,BanerjeeGhosh_Science_2018,Ghosh_JPCL_2020} However, despite concurrent advances in theoretical models to describe this effect,\cite{Dalum_NL_2019,yang_PRB_2019,Ghazaryan_JPCC_2020,Dianat_NL_2020,Zollner_JCTC_2020,Fransson_NL_2021,Franssin_JPCL_2022} questions remain regarding the role of spin coherence and the magnitude of spin polarization accompanying electron transport within chiral molecules. Moreover, the majority of CISS measurements have relied on ferromagnetic electrodes, which may perturb observed polarizations due to the interface between ferromagnetic surfaces and the chiral material.\cite{Waldeck_APLMaterials_2021} \cite{Liu_NatMater_2021,Alwan_JACS_2021,Vittmann_JPCL_2022} 

Recent proposals outline methods for studying CISS \textit{via} magnetic-resonance experiments.\cite{Chiesa_JPCL_2021,Fay_JPCL_2021,Luo_NJP_2021} Chiesa \textit{et al.} described methods for detection through either asymmetry in the dipolar coupling between the sensor and the polarized spin or \textit{via} polarization transfer from the polarized spin to the sensor.\cite{Chiesa_JPCL_2021} Meanwhile, Fay \textit{et al.} suggested monitoring coherence associated with CISS in the presence of a magnetic field that is orthogonal to the chiral axis.\cite{Fay_JPCL_2021} Critically, proposed experiments necessitate well-aligned molecular assemblies. Moreover, application of conventional NMR and EPR measurements may be limited by the need for substantial magnetic fields and large spin ensembles. In contrast, NV centers present a promising method for detection of CISS locally at the nanoscale. Notably, NVs were used to indirectly probe CISS by Merzeida \textit{et al},\cite{Meirzada_ACSNano_2021} who measured the magnetization orientation in a ferromagnetic layer that exhibited perpendicular anisotropy from chiral molecular adsorption. In future experiments, chiral molecules could be instead assembled directly on diamond surfaces, in close proximity to NV centers, with externally driven charge transfer to generate spin polarization with a chiral bridge.\cite{Abendroth_ACSNano_2017,Abendroth_ACSNano_2019,Olshansky_JACS_2019,Junge_JOC_2020,Lorenzo_JACS_2021} Adsorbed molecules supporting charge- and spin-polarized electrons would induce both electric and magnetic fields at the NV position, enabling new methods for characterization of CISS that could provide novel mechanistic insight. 

\subsection{Catalysis}
As previously mentioned, solid-state NMR is a powerful tool for studying the structure of heterogeneous catalysts\cite{Coperet_JAmChemSoc_2017} due to advancements in magnetic-field strength, \cite{Callon_JBiomolNMR_2021} fast MAS,\cite{Boeckmann_JMagnReson_2015} and dynamic nuclear polarization (DNP).\cite{Corzilius_2020_AnnuRevPhysChem}
However, it is often unclear whether the relevant, active species are observed since the working conditions of most catalysts include either high temperatures or pressures, which differ dramatically from standard measurement conditions.\cite{Yakimov_JPCL_2020} Consequently, the development of techniques for variable temperature and pressure operando NMR\cite{Jaegers_AccChemRes_2020} are viewed as an important step towards \textit{in situ} characterization of catalytic materials. However, such schemes are still limited to probing sensitive NMR nuclei that are in high abundance. Moreover, the detection of active sites in supported single-site catalysts is not compatible with this technique since such studies rely on solid-effect DNP and therefore must be performed at liquid-nitrogen temperatures in frozen glassy matrices of radical solution.\cite{Rossini_2013_AccChemRes} In contrast, the high pressure and temperature stability of diamond, in combination with exquisitely sensitive NV sensors, could enable detection of dilute, insensitive nuclear species at relevant catalytic working conditions. For such experiments, single-site catalysts would be prepared in the vicinity of NV sensor spins using a surface-organometallic chemistry.\cite{Coperet_2003_Angewandte} The conventional metal-oxide support (\textit{e.g.}, SiO$_2$, Al$_2$O$_3$) could be replaced either by well-defined diamond surface terminations with controllable densities of anchoring sites (\textit{e.g.}, hydroxyls or amines) or by a thin layer of oxide\cite{Liu_PNAS_2022}. The latter option would utilize the original grafting chemistry at the cost of larger NV-sample distances, as exemplified in Fig.\ref{fig:applications} for selected heterogeneous olefin-metathesis catalysts.\cite{Coperet_2003_Angewandte} Subsequently, structural changes resulting from catalytic reactions (\textit{e.g.}, loss of a ligand or change in oxidation state of the metal) would be detected using the NV center. Notably, while the large chemical-shift range and quadrupolar nature of many interesting NMR-active metals (\textit{e.g.}, $^{195}$Pt, $^{71}$Ga) are a major challenge in conventional NMR spectroscopy, they could serve as a resource for NV-center sensing. 

\section{Outlook}

NMR and EPR spectroscopy of molecules are powerful tools for elucidating details of chemical structure, intermolecular interactions, and chemical reactions. The experimental approach wherein traditional inductive detectors are replaced with NV centers in diamond has facilitated highly sensitive and localized spectroscopy, allowing for detection of electron and nuclear spins in external targets at the single- or few-spin level. Encouragingly, vast improvements in measurement sensitivity have been obtained through advances in diamond surface engineering that reduce charge instability and decoherence for shallow emitters. In addition, novel methods for chemical functionalization of diamond allows for deterministic, covalent attachment of target molecular species. Despite this impressive progress, further experimental advances are required to achieve coherent detection of single nuclear spins on the diamond surface. 

Additional experimental challenges must also be overcome. For instance, the stability of molecular films or of hierarchically ordered structures in biomolecules may be compromised by extended exposure to laser- and microwave-induced heating. Nevertheless, use of low laser power as well as reduced experimental duty cycles may help to prevent ablation of specifically bound molecules during extended measurement times. Likewise, reduced microwave duty cycles, or elimination of microwave sources entirely (\textit{e.g.}, for relaxometry measurements) can mitigate heating issues. In addition, current detection sensitivities necessitate long integration times, placing limits on reaction monitoring or tracking of transient paramagnetic intermediates. Greater control over molecular density, orientation, and binding configurations at diamond surfaces will circumvent rapid diffusion of analytes into and out of NV sensing volumes, allowing for longer measurements. Moreover, motional dynamics and thermal fluctuations in large molecules induce signal broadening, complicating the extraction of structural and functional information and localized imaging of surface spins. Depending on the sensing application, immobilization approaches that restrict motional degrees of freedom can be employed to reduce spectral broadening. Finally, despite the exquisite sensitivity and spatial resolution, NV-center spectroscopy cannot achieve the spectral resolution of conventional spectroscopy methods. Indeed, typical magnetic-field biases applied in NV experiments are on the order of \textit{ca.} $10^{-3}-10^{-1}$ T, compared to \textit{ca.} $10^0-10^1$ T in conventional NMR. Therefore, a 5 ppm $^1$H chemical shift, easily resolved in conventional NMR with an operating field strength of, \textit{e.g.}, 600 MHz, would result in an absolute shift of $<1$ Hz at 20 mT. However, novel NV approaches including homonuclear decoupling sequences,\cite{Aslam_Science_2017} ensemble measurements,\cite{Glenn_Nature_2018} or hyperpolarization techniques\cite{Arunkumar_PRXQuant_2021} provide promising paths forward.

In conclusion, advances in diamond surface engineering and chemical functionalization have facilitated landmark advances in nanoscale NMR and EPR using NV centers. Moving forward, the simultaneous immobilization of chemical analytes and preservation of NV sensing properties represents a critical step toward single-spin, coherent detection in chemically and biologically relevant systems. Efforts to overcome these challenges will benefit from the interdisciplinary nature of this research field, which lies at the intersection of surface chemistry, materials science, and quantum metrology. Motivated by the promise of new insights into
molecular structure, spin transport, and reaction monitoring, the utility of NV-center sensing for the chemical sciences has only just begun to be realized.

\section*{Author Contributions}
Following CRediT definitions: conceptualization: E. J., K. H., L. A. V., W. S. H., C. L. D., J. M. A.; funding acquisition: C. L. D., J. M. A.; investigation: E. J., K. H., L. A. V., W. S. H., J. M. A.; project administration: J. M. A.; supervision: J. M. A.; validation: E. J., K. H., L. A. V., W. S. H., C. L. D., J. M. A.; visualization: E. J., K. H., L. A. V., J. M. A.; writing -- original draft: E. J., K. H., L. A. V., W. S. H., J. M. A.; writing -- review \& editing: E. J., K. H., L. A. V., W. S. H., C. L. D., J. M. A.

\section*{Conflicts of interest}
There are no conflicts to declare.

\section*{Acknowledgements}
The authors thank Lilian Childress, Claire McLellan, Marius Palm, Romana Schirhagl, and Pol Welter for helpful comments and discussions. We gratefully acknowledge the Swiss National Science Foundation (SNSF) Ambizione Grant, "Quantum Coherence in Mirror-Image Molecules" [PZ00P2\_201590], SNSF Project Grant No. 200020 175600, and funding through the SNSF NCCR QSIT, a National Centre of Competence in Research in Quantum Science and Technology, Grant No. 51NF40-185902 for support of this work. We also gratefully acknowledge the European Research Council through ERC CoG 817720 (IMAGINE) and the Advancing Science and TEchnology thRough dIamond Quantum Sensing (ASTERQIS) program, Grant No. 820394, of the European Commission. J. M. A. also acknowledges an ETH Career Seed Grant.

\balance

\inputencoding{utf8}
\bibliography{rsc} %
\bibliographystyle{rsc} %

\end{document}